\documentclass[11pt]{article}
\usepackage{a4}
\usepackage{latexsym}
\usepackage{marvosym}
\usepackage{graphicx}
\usepackage{amssymb}
\usepackage{xspace}
\usepackage{cite}
\usepackage{url}
\usepackage{float}
\usepackage{algorithmicx}
\usepackage[ruled]{algorithm}
\usepackage{algpseudocode}
\usepackage{flafter}

\newbox\tmptwocolbox
\long\def\twocol#1{\setbox\tmptwocolbox=\vtop{\hbox{}\hsize=0.5\hsize\textwidth=\hsize\columnwidth=\hsize #1\par}%
\hbox to\hsize{\hfill\hbox to 0.49\hsize{\vtop{\hbox{}\hsize=0.49\hsize\vsplit\tmptwocolbox to 0.5\dp\tmptwocolbox}\hss}\hfill
  \hbox to 0.49\hsize{\vtop{\hbox{}\hsize=0.49\hsize\copy\tmptwocolbox}\hss}\hfill}}

\newtheorem{definition}{Definition}
\newtheorem{theorem}{Theorem}
\newtheorem{lemma}{Lemma}
\newtheorem{corollary}{Corollary}

\newenvironment{proof}{{\em Proof:}~ }{\nobreak\par\nobreak\hfill$\Box$ \vskip 2mm}
\newenvironment{proofsk}{{\em Sketch of proof:}~ }{\nobreak\par\nobreak\hfill$\Box$ \vskip 2mm}
\newenvironment{proofof}[1]{\def\proofofnametmp{#1}\begin{proofoftmp}}{\end{proofoftmp}}
\newenvironment{proofoftmp}{{\em Proof of {\proofofnametmp}}}{\nobreak\par\nobreak\hfill$\Box$ \vskip 2mm}
\newtheorem{obs}{Observation}

\newcommand{\bm}[1]{{\mbox{\boldmath $#1$}}}

\newcommand{\myfig}[3]{\begin{figure}[h]\centerline{\includegraphics[width=#2]{fig#1.eps}}\caption{#3}\label{fig:#1}\end{figure}}

\newcommand{\lla}{\left\langle}
\newcommand{\rra}{\right\rangle}

\newcommand{\Ia}{{\bf P1}\xspace}
\newcommand{\Ib}{{\bf P2}\xspace}
\newcommand{\Ic}{{\bf P3}\xspace}
\newcommand{\Id}{{\bf P4}\xspace}

\begin{document}
\floatplacement{algorithm}{htb} \floatplacement{figure}{htb}

\title{Online Bandwidth Allocation}
\author{
Michal Fori\v{s}ek\and Branislav Katreniak\and Jana Katreniakov\'a
\and Rastislav Kr\'alovi\v{c}\and Richard Kr\'alovi\v{c}\and Vladim\'ir Koutn\'y
\and Dana Pardubsk\'a\and Tom\'a\v{s} Plachetka\and Branislav Rovan \\[2mm]
 Dept. of Computer Science, Comenius University,\\
 Mlynsk\'a dolina, 84248 Bratislava, Slovakia.
}

\maketitle

\begin{abstract}
The paper investigates a version of the resource allocation problem arising in
the wireless networking, namely in the OVSF code reallocation process.  In this
setting a complete binary tree of a given height $n$ is considered, together
with a sequence of requests which have to be served in an online manner.  The
requests are of two types: an insertion request requires to allocate a complete
subtree of a given height, and a deletion request frees a given allocated
subtree. In order to serve an insertion request it might be necessary to move
some already allocated subtrees to other locations in order to free a large enough subtree. 
We are interested in the worst case average number of such reallocations needed to serve a
request.

In \cite{E+04} the authors delivered bounds on the competitive ratio of online
algorithm solving this problem, and showed that the ratio is between $1.5$ and $O(n)$.
We partially answer their question about the exact value by giving an $O(1)$-competitive 
online algorithm.

In \cite{DCC05}, authors use the same model in the context of memory management systems,
and analyze the number of reallocations needed to serve a request in the worst case. In this setting,
our result is a corresponding amortized analysis.

\vskip 3mm
\noindent
{\bf Classification:} Algorithms and data structures
\end{abstract}

\section{Introduction and motivation}

Universal Mobile Telecommunications System (UMTS) is one of the
third-gen\-er\-a\-tion (3G) mobile phone technologies that uses W-CDMA as the
underlying standard, and is standardized by the 3GPP \cite{W}. 
The W-CDMA (Wideband Code Division Multiple
Access) is a wideband spread-spectrum 3G mobile telecommunication air interface
that utilizes code division multiple access.  The main idea behind the W-CDMA
is to use physical properties of interference: if two transmitted signals
at a point are in phase, they will "add up" to give twice the amplitude of each
signal, but if they are out of phase, they will "subtract" and give a signal
that is the difference of the amplitudes. Hence, the signal received by a
particular station is the sum (component-wise) of the respective transmitted
vectors of all senders in the area. In the W-CDMA, every sender $s$ is given a
{\em chip code} $\bm{v}$.  Let us represent the data to be sent by a vector of
$\pm1$.  When $s$ wants to send a {\em data vector} $\bm{d}=(d_1,\ldots,d_n)$,
$d_i\in\{1,-1\}$, it sends instead a sequence
$d_1\cdot\bm{v},d_2\cdot\bm{v},\ldots,d_n\cdot\bm{v}$, i.e.\ $n$-times the chip
code modified by the data.  For example, consider a
sender with a chip code $(1,-1)$ that wants to send data $(1,-1,1)$; then the
actually transmitted signal is $(1,-1,-1,1,1,-1)$.  The signal received by a
station is then a sum of all transmitted signals. Clearly, if the chip codes
are orthogonal, it is possible to uniquely decode all the signals. 

\myfig{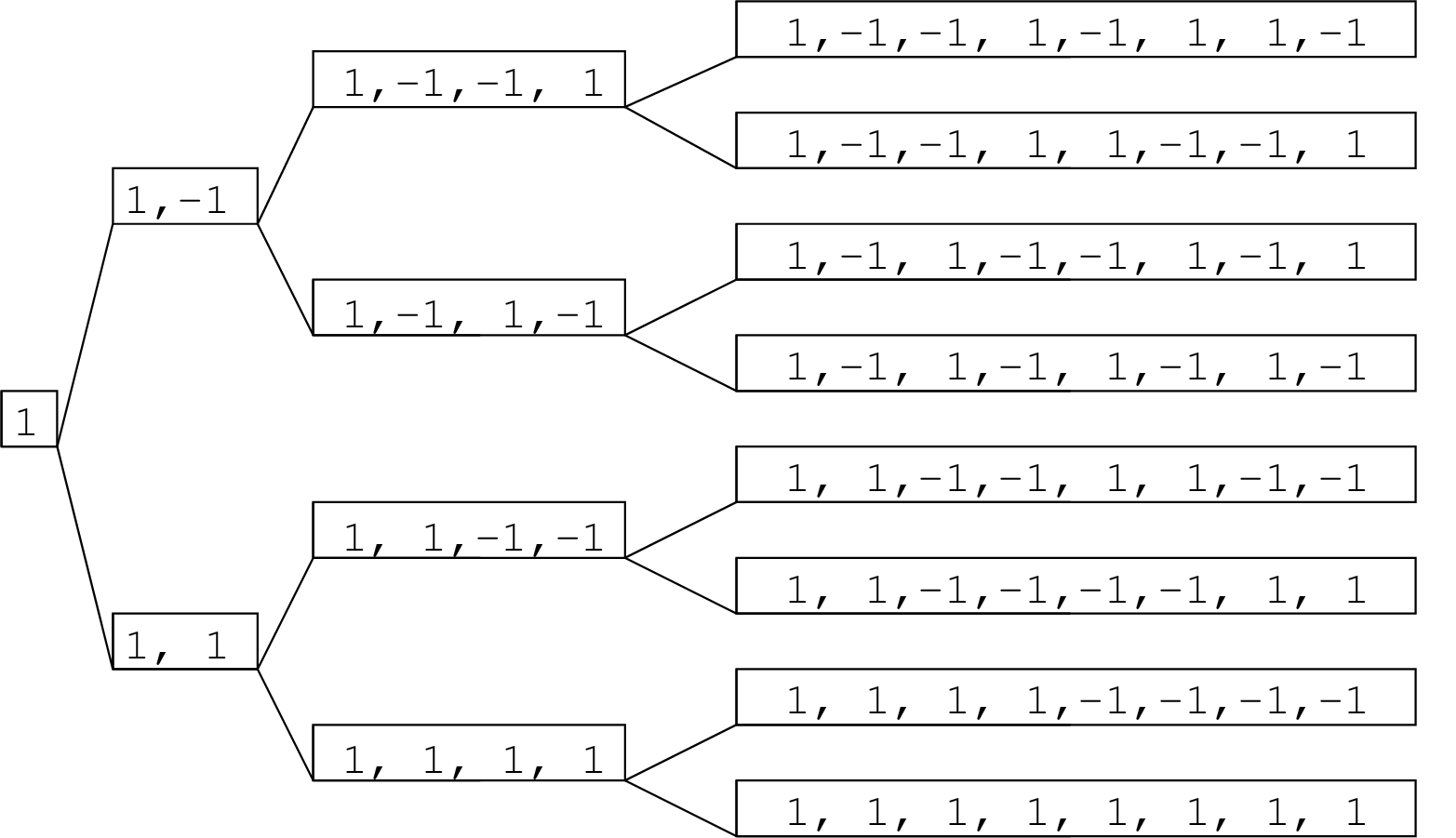}{8cm}{An OVSF tree}

One commonly used method of implementing the chip code allocation is Orthogonal
Variable Spreading Factor Codes (OVSF). Consider a complete binary tree, where
the root is labeled by $(1)$ , the left son of a vertex with label $\alpha$ is
labeled $(\alpha,\alpha)$ and the right son is labeled $(\alpha,-\alpha)$ (see
Figure~\ref{fig:001}). 

If a sender station enters the system, it is given a chip code from the tree in
such a way that there is at most one assigned code from each root-to-leaf path.
It can be shown \cite{MS00,ASO97} that this construction fulfills the orthogonality
property even with codes of different lengths. 

Clearly, a code at depth $l$ in the tree has length $2^l$, and a sender
using this code will use a fraction of $1/2^l$ of the overall bandwidth.  When
users enter the system, they request a code of a given length. It is irrelevant
which particular code is assigned to which user, the length is the only thing
that matters.
When users connect to and disconnect from a given base station, i.e.\ request and
release codes, the tree can become fragmented. It may happen that no code at
the requested level is available, even though there is enough bandwidth (see
Figure~\ref{fig:002}).

\myfig{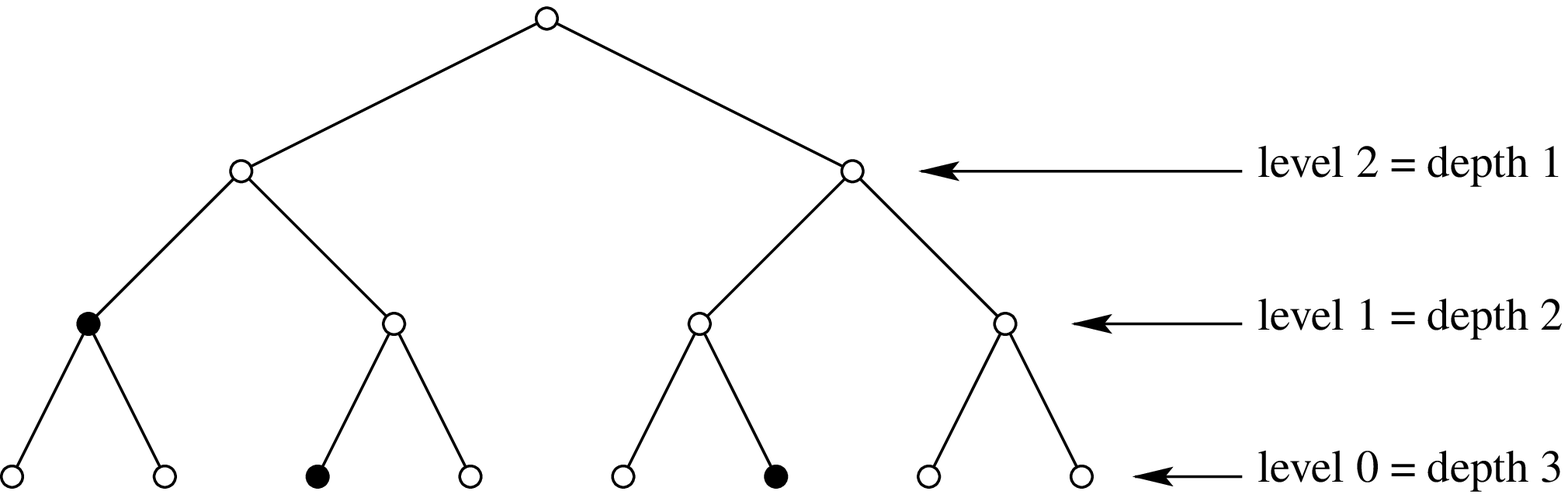}{9cm}{No code at depth 2 can be allocated although there is enough free bandwidth.
Full circles represent allocated vectors.}

This problem can be solved by changing the chip codes of some already
registered users, i.e.\ reallocating the vertices of the tree. Since the cost of
a reallocation dominates this operation, the number of reallocations should be
kept minimal.
In \cite{E+04} the authors considered the problem of minimizing the number of
reallocations over a given schedule and showed that it is NP-hard to generate
an optimal allocation schedule. In this paper, we show that the online version
proposed in \cite{E+04} can be solved in amortized complexity $O(1)$
reallocations per request. The technical parts can be found
in the Appendix.

\section{Problem definition}

Consider a complete binary tree $T=(V,E)$ of height $n$. Leaves are said to
be at level $0$, and the root at level $n$. A {\em request vector} is a vector
$\bm{r}=(r_0,\ldots,r_n)\in{\mathbb N}^{n+1}$, where $r_i$ represents the number of
users that request a code at level $i$. A {\em code assignment} of a particular
request vector \bm{r} is a subset of vertices $F\subset V$, such that every
path from a leaf to the root contains at most one vertex from $F$, and there
are exactly $r_i$ vertices at level $i$ in $F$. The input consists of a
sequence of requests of two types: {\em insertion} and
{\em deletion}. Suppose that, after $t$ requests the request vector is 
$\bm{r}_t$ and the corresponding code assignment is $F_t$. The algorithm has to process
the next request in the following way:

\begin{itemize}
\item[{\bf A)}] {\bf insertion request} at a vertex at level $i$.\\
The algorithm must output a new code assignment $F_{t+1}$ satisfying the
request vector $\bm{r}_{t+1}=(r_0,\ldots,r_i+1,\ldots,r_n)$\footnote{
We may consider, without loss of generality, that there is enough bandwidth to satisfy each request.
}
\item[{\bf B')}] {\bf deletion request} of a particular vertex $v\in F_t$.\\
Let $v$ be at level $i$.
The new request vector is $\bm{r}_{t+1}=(r_0,\ldots,r_i-1,\ldots,r_n)$, and the new
code assignment is $F_{t+1}=F_t\setminus\{v\}$.
\end{itemize}

During each step, the number of reassignments is $|F_{t+1}\setminus F_t|$. For a
given sequence of requests $R_1,\ldots,R_m$, we are  interested in the
amortized number of reassignments per request, i.e.\ the quantity
$\frac{1}{m}\displaystyle\sum_{t=0}^{m-1}|F_{t+1}\setminus F_t|$

First, let us note that there is no harm in allowing reallocations also in the
deletion requests: in a deletion request the algorithm remembers the moves it
would perform, and performs them in the next insertion request\footnote{In case
of consecutive deletion requests the algorithm can clearly
maintain a mapping between the actual vertices in $F$ and their "virtual"
positions.}.  As a next step, when deleting a particular vertex $v$ at level
$i$, the algorithm may delete any other vertex at the same level and then
reallocate the code of the deleted vertex to $v$. These arguments allow us to
reformulate the requirements for processing deletion requests as
follows:

\begin{itemize}
\item[{\bf B)}] {\bf deletion request} of a vertex at level $i$.\\
The algorithm must output a new code assignment $F_{t+1}$ satisfying the
request vector $\bm{r}_{t+1}=(r_0,\ldots,r_i-1,\ldots,r_n)$.
\end{itemize}

This definition bears resemblance to memory allocation problems studied in the
operating systems community, in particular to the {\em binary buddy
system} memory allocation strategy introduced in \cite{K65}. In
this strategy, requests to allocate and deallocate memory blocks of sizes
$2^l$ are served. The system maintains a list of free blocks of sizes of
$2^k$. When allocating a block of size $2^l$ in a situation where no
block of this size is free, some bigger block is recursively split into
two halves called {\em buddies}. When a block whose buddy is free is
deallocated, both buddies are recursively recombined into a bigger block.

The properties of binary buddy system have been extensively studied in the
literature (see e.g. \cite{BDM05} and references therein). However, the
bulk of this research is focused on cases without reallocation of memory
blocks. E.g. \cite{BDM05} shows how to implement the buddy system in
amortized constant time per allocation/deallocation request, but in a model
where reallocation is not allowed.

A binary buddy system with memory block reallocations has been studied in
\cite{DCC05}. This paper analyzes both the number of reallocated blocks and the
number of reallocated bytes per request; the analysis of the number of
reallocated block is in fact the very same model that is used in our paper.
However, only the worst case scenario is dealt with in \cite{DCC05}, hence
our results are relevant also to the recent memory allocation research.

\section{The algorithm}

We propose an online algorithm that processes the sequence of requests
according to the rules A) and B).  If we order the leaves from left to right, and
label them with numbers $0,\ldots,2^n-1$, there is an interval of the form
$I_u=\lla i2^l,(i+1)2^l-1\rra$ assigned to each tree vertex $u$ of level $l$. We
shall call each such interval a {\em place} of level $l$, and we say that $I_u$
begins at position $i2^l$.  For a given code assignment $F$, we shall call a
place $I_u$ corresponding to a vertex $u$ an {\em empty} (or {\em free}) {\em place} if neither
$u$ nor any vertex from the subtree rooted at $u$ is in $F$.  If $u\in F$ the
corresponding place $I_u$ is an {\em occupied place}, and we say that there is a {\em pebble} of level $l$ located on
$I_u$ (or, alternatively there is a pebble of level $l$ at position $i2^l$).  Since
in a code assignment $F$, every path from a leaf to the root contains at most
one vertex, we can view the code assignment $F$ as a sequence of disjoint
places which are either empty or occupied by pebble (see Figure~\ref{fig:003}). While the free places
in this decomposition are not uniquely defined, we shall overlook this
ambiguity, as we will argue either about a particular place or about the
overall size of free places (called also {\em free bandwidth}).

We shall say that a pebble (or place) of level $l$ has {\em size} $2^l$. 
The left (right) neighbor of a pebble is a pebble placed on the next occupied place to the left (right)\footnote{
Sometimes we talk about a left (right) neighbor of a free place.}.
We shall denote pebbles by capital letters $A,B,\ldots,X$, and their
corresponding sizes by $a,b,\ldots,x$. Sometimes we shall use the notion of
a vertex and the corresponding place interchangeably.

\myfig{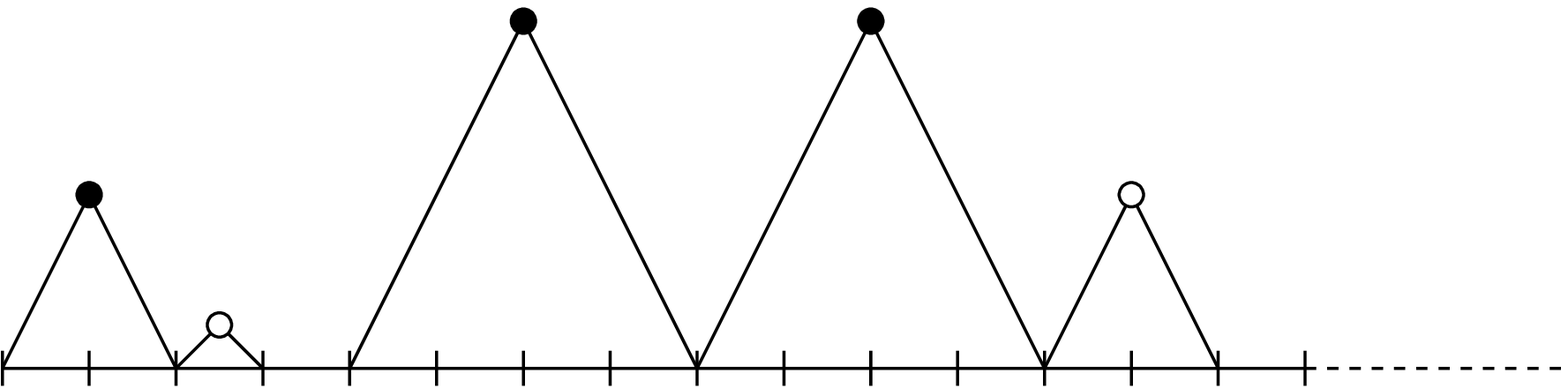}{8cm}{A code assignment seen as a sequence of black and white pebbles.}

The key idea of our algorithm is to maintain a well defined structure in the sequence of pebbles.
An obvious approach would be to keep the sequence sorted -- i.e. the pebbles of lower levels
always preceding the pebbles of higher levels with only the smallest necessary free places
between them. It is easy to see that the addition and deletion to such structured sequence
can be done with at most $O(n)$ reallocations.\footnote{
The addition request is processed by placing the pebble of level $l$ at the end of sequence of pebbles of
this level. If this place is already occupied by some pebble $B$, $B$ is removed and reinserted. Since there
are at most $n$ different levels, and the levels of reinserted pebbles are increasing, the whole process
ends after $O(n)$ iterations. The deletion works in a similar fashion.}
Unfortunately, it is also not difficult to see that there is a sequence of requests such that
this approach needs amortized $O(n)$ reallocations per request (see
\cite{E+04}). The problem is that a too strictly defined
structure needs too much "housekeeping" operations.

The structure maintained by our algorithm will be less rigid. Basically, we will maintain a sorted
sequence as in the above example, however, not all pebbles must be included in this sequence. Pebbles that form a
sorted sequence according to the previous rule will be called {\em black} pebbles. There can be also {\em white}
pebbles present that do not fit into this structure but we impose other restrictions on them.
Informally, if there is a free place in the black sequence, a single white pebble can be placed at the
beginning of this place. Moreover, all white pebbles form an increasing sequence.
Formally, the structure of the sequence of pebbles is described by the following invariants. It can be shown
that they indeed imply the informal description above.

\begin{itemize}
\item[{\bf C:}] Pebble $A$ is {\em black}, if there is no bigger pebble before $A$, otherwise it is {\em white}\\[-2ex]
\item[\Ia:] The free bandwidth before any pebble $X$ is strictly less than $x$.
\item[\Ib:] There is always at least one black pebble between any two white pebbles.
\item[\Ic:] There is no white pebble such that both its left and right neighbor are black and of the same size. 
\end{itemize}

A sequence of pebbles and free places satisfying \Ia--\Ic will be called a {\em valid situation}. The key observation
about valid situations is that in a valid situation it is always possible to process an insertion request
without reallocations:

\begin{lemma}\label{lemma:canPut}
Consider a valid situation, and an insertion request of size $a=2^l$.  If there is a free
bandwidth of at least $a$, then there is also a free place of size $a$.
\end{lemma}

Our algorithm maintains valid situations over the whole computation, so all requests can
be processed without reallocations. However, processing a request may result in a situation that
is not valid. The most difficult task is to develop a post-processing phase in each request that
restores the validity using only few reallocations. We show that in the case of insertion requests,
a constant number of reallocations in each request is sufficient. In the deletion requests, however,
a more involved accounting argument is used to show that the {\em average} number of reallocations
per request in a worst-case execution remains constant.

\subsection{Procedure {\sc Insert}}

Before presenting the procedure for managing insertion requests, we need an additional definition.

\begin{definition}
The {\em closing position} of level $l$ (of size $x$) is the position after
the last black pebble of level $l$ (of size $x$) if such pebble exists.
Otherwise, the closing position of level $l$ is the position of the first pebble
of level bigger than $l$ (size bigger than $x$).
\end{definition}

\begin{algorithm}
\caption{Procedure {\sc Insert}: inserts a pebble $A$ of size $a$ into a valid situation.}
\label{alg:insert}
\twocol{
\begin{algorithmic}[1]
{\small
\State\label{alg:insert.1}let $P$ be the first free place of size $a$
\State let $B$ be the left neighbor of $P$
\If {{\em $B$ does not exist or $B$ is black\hfil\break}}
  \State\label{alg:insert.2}put $A$ on $P$
  \State {\tt return}
\EndIf
\State
\State\label{alg:insert.3}let $C$ be the left neighbor of $B$
\If {{\em $c<a$}}
  \State\label{alg:insert.4}put $A$ on $P$
  \State {\tt return}
\ElsIf {{\em $c=a$}}
  \State\label{alg:insert.5}remove $B$
  \State put $A$ just after $C$
  \State put $B$ just after $A$
  \State\label{alg:insert.6}{\tt return}
\EndIf
\State
\State remove $B$
\If {$a<b$}
  \State rename $A$ and $B$ so that $A$ \hfil\break\hbox{}\ \ \ \ \ is the bigger pebble
\EndIf
\State
\State let $D$ be the pebble at the closing \hfil\break position of size $a$.
\If {{\em $D$ is white}}
  \State\label{alg:insert.7} let $E$ be $D$'s right neighbor
  \State remove $D$, $E$
  \State put $A$ at $D$'s original position
  \State put $B$ after $A$
  \State put $D$ after $B$ 
  \State\label{alg:insert.8} put $E$ at $B$'s original position
\ElsIf {{\em $D$ is black}}
  \State\label{alg:insert.9} remove $D$
  \State put $A$ at $D$'s original position
  \State\label{alg:insert.10} put $B$ after $A$, put $D$ after $C$
\EndIf
}
\end{algorithmic}
}
\end{algorithm}

Procedure {\sc Insert} processes a new request of size $a$ in a valid situation. First, a free place
of size $a$ is found, and a pebble is put
on this place. If the sequence is no longer valid after this operation, the algorithm reassigns a
constant number of pebbles in order to restore the invariants. The procedure is
listed as Algorithm~\ref{alg:insert}, and its analysis is given in the following theorem:

\begin{theorem}\label{thm:insertCorrect}
Consider a sequence of pebbles and free places that forms a valid situation,
and an insertion request of size $a$.  If there is enough bandwidth,
procedure {\sc Insert} correctly processes the request. Moreover, after
finishing, the situation remains valid, and only a constant number of pebbles
has been reassigned.
\end{theorem}
\begin{proofsk}
Here, we present an overall structure of the proof with a number of unproven claims.
The complete version can be found in Appendix.

Consider an insertion request of size $a$, and suppose the invariants hold.
Because of Lemma~\ref{lemma:canPut} the line \ref{alg:insert.1} of Algorithm
\ref{alg:insert} is correct, i.e.\ there is a free place of size $a$. Let $P$
be the first such free place.
If $P$ does not have a left neighbor (i.e. there is no other pebble present), 
the algorithm puts $A$ on $P$ in line
\ref{alg:insert.2} and exits.
From now on suppose that $P$ has a left neighbor $B$, and denote the potential
right neighbor of $P$ by $Z$.  The rest of the proof consists of a case
analysis of a number of cases as they follow from Algorithm~\ref{alg:insert}.  
For each case, the action of the algorithm is analyzed and it is proven that the
resulting situation is valid.

The easy part is when $B$ is black, since in this case the algorithm puts $A$ on $P$ 
in line \ref{alg:insert.2} and exits. If $B$ is white, however, there exists a left neighbor $C$ of
$B$, such that $C$ is black and $c>b$.  
We distinguish three sub-cases: $c<a$, $c=a$, and $c>a$. The case $c<a$
is handled by putting $A$ on $P$ in line~\ref{alg:insert.4}. If $c=a$,
the sequence of lines \ref{alg:insert.5}--\ref{alg:insert.6} is executed:
first, pebble $B$ is temporarily
removed. Since there has been no other pebble between $A$ and $C$, and $a=c$,
the place of size $a$ immediately following $C$ is now free.  Put $A$
immediately after $C$. Since $b<c=a$, the place of size $b$ immediately
following $A$ is now free, so $B$ can be put there. Finally, if $c>a$,
the algorithm temporarily removes $B$\footnote{assume w.l.o.g. that $a\ge b$, 
see Appendix}, and calls the pebble at the closing position of size
$a$ by $D$ (there must be a pebble present). The proof is concluded be considering two
final sub-cases based on whether $D$ is white or black. In the first case, the action is
depicted on Figure~\ref{fig:010}.

\myfig{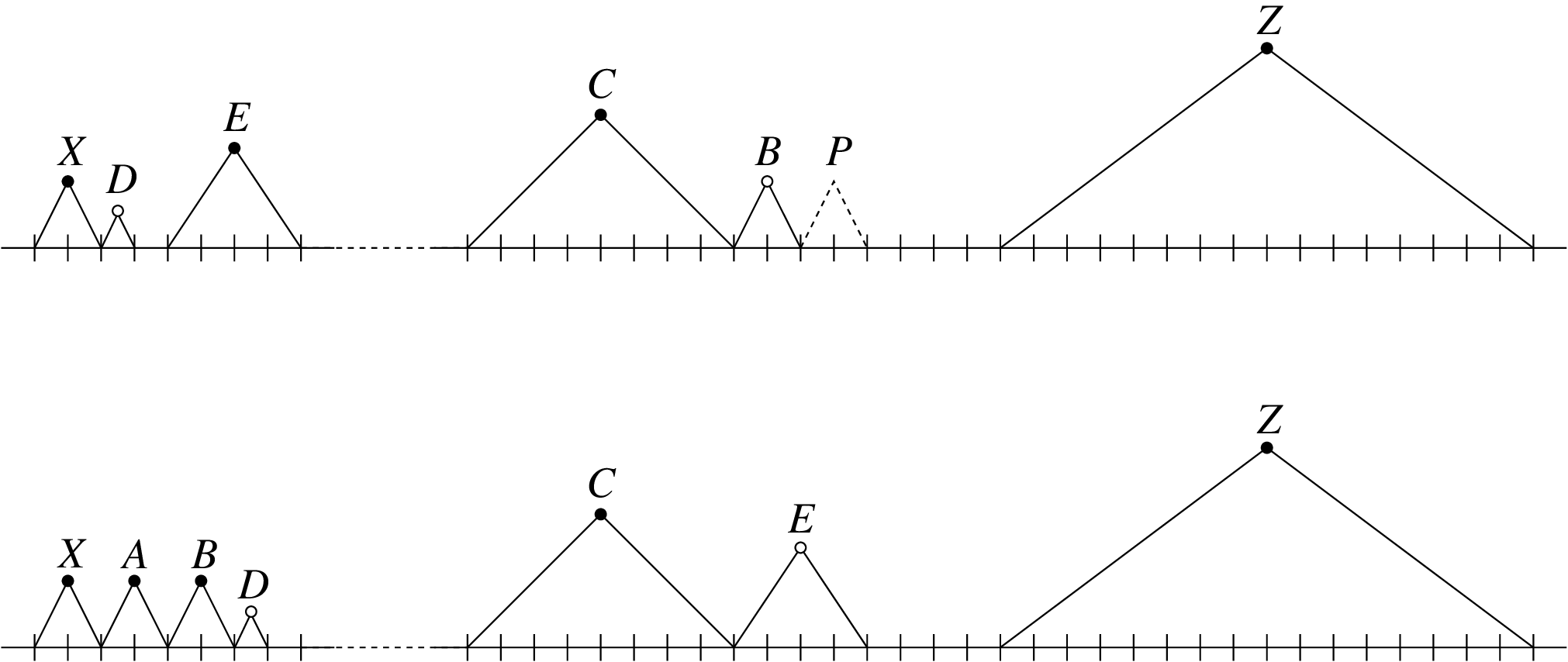}{12cm}{An example of executing lines
\ref{alg:insert.7}--\ref{alg:insert.8} of procedure {\sc Insert}}

If $D$ is black, the situation is as follows from Figure~\ref{fig:012}.

\myfig{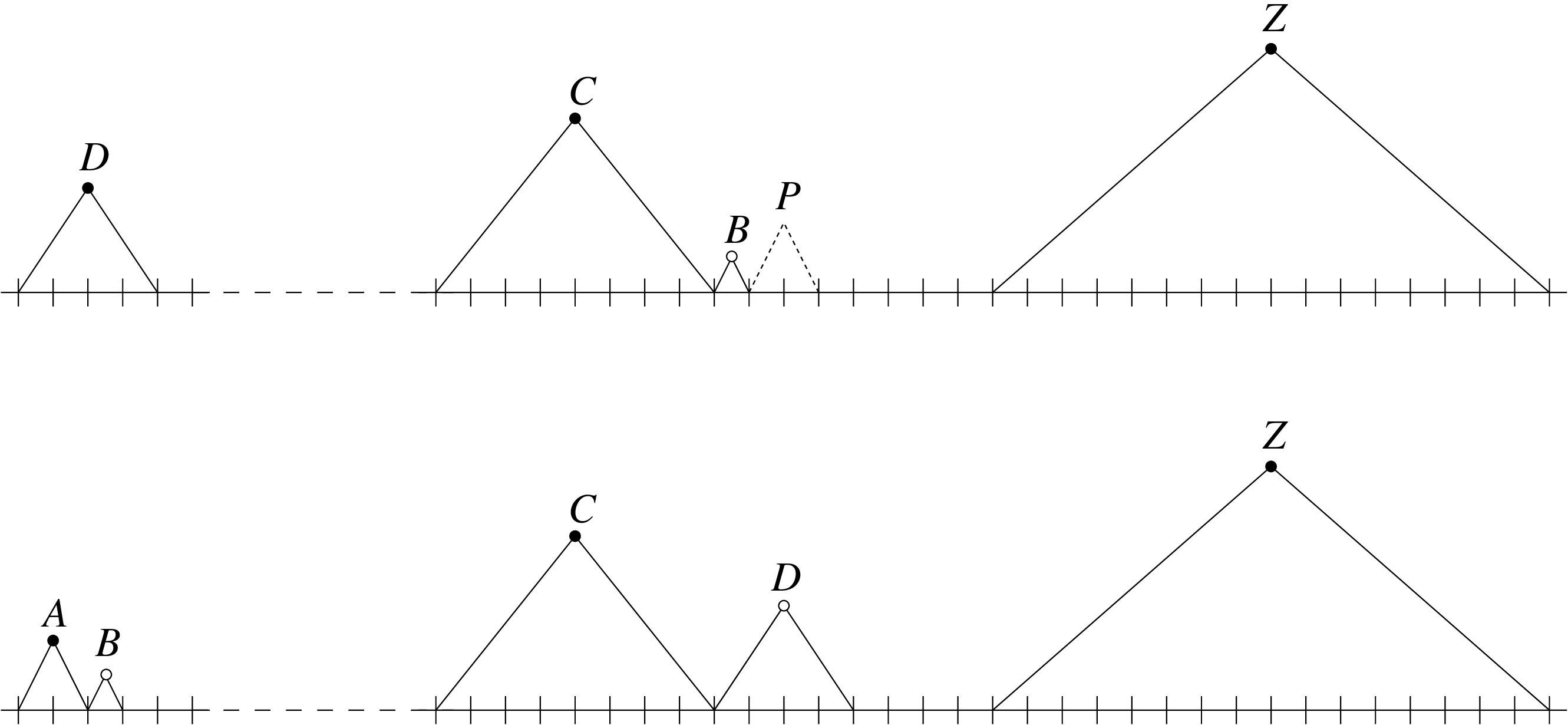}{12cm}{An example of executing lines
\ref{alg:insert.9}--\ref{alg:insert.10} of procedure {\sc Insert}}
\end{proofsk}

\subsection{Procedure {\sc Delete}}

\begin{algorithm}
\caption{Procedure {\sc Delete} that removes the last pebble of level $l$.}
\label{alg:delete}
\begin{algorithmic}[1]
{\small
\State let $A$ be the last pebble of level $l$, and $i$ be the starting position of $A$
\State remove $A$
\While {{\em there are any pebbles to the right of $i$}}
  \State let $x$ be the size of the smallest pebble to the right of $i$
  \If {{\em there is a free place of size $x$ starting at $i$}}
    \State\label{alg:delete.1}let $X$ be the rightmost pebble of size $x$
    \State let $j$ be the starting position of $X$
    \State move $X$ to $i$
    \If {{\em $X$ has white left neighbor $Q$, and $Q$ has a left neighbor $W$ of size $x$}}
      \State\label{alg:delete.2}swap $X$ and $Q$
    \EndIf
    \State let $i$:=$j$
  \Else
    \State {\tt exit}
  \EndIf
\EndWhile
}
\end{algorithmic}
\end{algorithm}

The deletion request requires to remove one pebble of a specified level.
Procedure {\sc Delete} (see Algorithm~\ref{alg:delete}) first removes the last
(rightmost) pebble $A$ of the requested level. However, it may happen that this
action violates the invariants \Ia--\Ic. To remedy this, the algorithm uses
several iterations to ``push the problem'' to the right.
The ``problem'' in
this case is in the free place caused by removing $A$. 
The ``pushing'' is done by selecting a suitable pebble $X$ to the right of $A$,
removing it, and using it to fill in the gap. A new iteration then starts to fix
the problem at $X$'s original place. The suitable candidate is found as follows:
from among all pebbles to the right of $A$, select the smallest one. If there
are more pebbles of the smallest size, select the rightmost one. To argue the correctness
it is needed to show that this procedure is well defined, and that after a finite number of iterations, a valid situation
is obtained (the full proof can be 
found in Appendix):

\begin{theorem}
\label{theorem:deleteCorrect}
Consider a sequence of pebbles and free places that forms a valid situation,
and a deletion request of size $a$.  Procedure {\sc Delete} correctly
processes the request. Moreover, after finishing, the situation remains
valid.
\end{theorem}
\begin{proofsk}
Let us number the iterations of the {\bf while} loop by $t=1,2,\ldots$.  Let
$\Gamma_t$ be the configuration of pebbles and free places at the beginning of
the $t$-th iteration.  Hence, the $t$-th iteration starts with $\Gamma_t$, and
a position $i_t$; it selects a pebble $X_t$ of size $x_t$ starting at $j_t$,
moves it to $i_t$, and sets $i_{t+1}:=j_t$. Moreover, at the beginning of the
$t$-th iteration we shall consider a free place $P_t$ of size $a_t$ starting at
$i_t$, such that $a_1=a$, and the size $a_{t+1}$ is determined as follows: let
$\Gamma'_t$ be obtained from $\Gamma_t$ by putting a new pebble $A_t$ on $P_t$.
If the color of $X_t$ in  $\Gamma'_t$ is black then let $a_{t+1}=x_t$,
otherwise let $Y_t$ be $X_t$'s left neighbor, and $a_{t+1}=y_t$. 
It can be shown that this definition is correct, i.e. that $P_{t+1}$ is indeed free.

\myfig{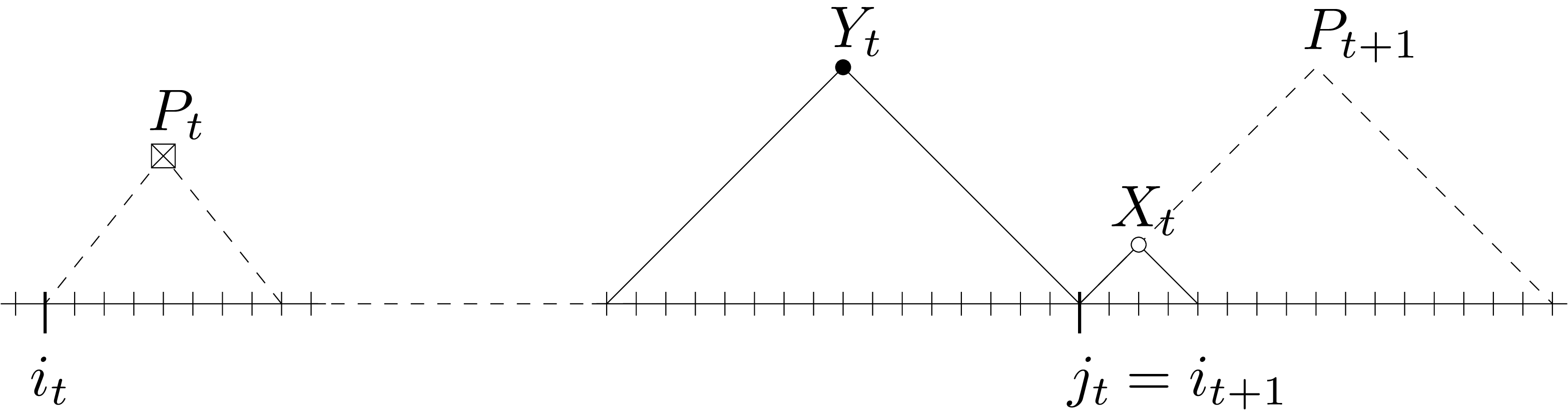}{11cm}{One iteration of the {\bf while} loop in procedure {\sc Delete}.}

\noindent
We shall prove the following claim by induction on $t$: 

\vskip 2mm
\noindent
{\em
For every iteration $t$ of the {\bf while} loop, the corresponding $\Gamma'_t$ is a
valid situation.
}

\vskip 2mm
\noindent
The first iteration starts after removing $A$ from a valid situation, so the
claim for $t=1$ holds.

\vskip 2mm
\noindent
Consider the $t$-th iteration. The algorithm either stops or enters the next
iteration.  We prove that in the latter case $\Gamma_{t+1}'$ is valid, provided
$\Gamma_t'$ was valid. The proof again continues with a case-analysis. First, if
$x_t<a_t$, it is possible to prove that there are neither white
pebbles nor free places between $i_t$ and $j_t$, from which the invariants readily follow.
On the other hand, if $x_t\ge a_t$, then the swap on line \ref{alg:delete.2} never
happens, and it is again possible to argue the validity of the resulting situation.

Having proved the claim, the proof is concluded by analyzing the last iteration:
the algorithm can stop either when there are no pebbles to the right of $i_t$,
or when the
selected pebble $X_t$ does not fit to position $i_t$. In both cases it is possible to 
show that the resulting situation is valid.
\end{proofsk}

\section{Complexity}

So far we have argued about the correctness of the algorithm, showing that
it correctly processes all requests. This section is devoted to the analysis of
the average number of reassignments per request needed in the worst-case computation.
Obviously, the only situation in which a non-constant number of reallocation could be performed
is the iteration of the {\bf while} loop in {\sc Delete}. Hence, our aim 
is to develop an accounting scheme that would ensure a linear (in the number
of requests) number
of iterations of the {\bf while} loop over the whole computation. To this end we 
introduce
the notion of {\em coins}: each request has associated a constant number of
coins which can be put on some places. Every iteration of the main loop in {\sc
Delete} consumes a coin. In the following we show where to put the constant
number of coins in every request such that each iteration of the loop in {\sc
Delete} can be paid by an existing coin.

In our coin placement strategy we shall maintain the following additional invariant:

\begin{itemize}
\item[\Id:] Consider a free place $P$ such that there are some pebbles to the right of $P$. 
Let $X$ be the smallest pebble to the right of $P$. Then there are
at least $\left\lfloor 2p/x\right\rfloor$ coins on $P$.
\end{itemize}

From now on we shall consider situations with some coins placed in some places, and we show how
to manage the coins so that there is always enough cash to pay for each iteration in {\sc Delete}.
The following two lemmas present the accounting strategy for {\sc Insert} and {\sc Delete}:

\begin{lemma}
Let us suppose that procedure {\sc Insert} was called from a situation
in which invariants \Ia -- \Id hold.
Then it is possible to add a constant number of coins and reallocate the existing ones
in such a way that invariants \Ia -- \Id remain valid.
\end{lemma}

\begin{proof} It has already been proven that invariants \Ia -- \Ic are
preserved by procedure {\sc Insert}, so it is sufficient to show how to add a
constant number of coins in order to satisfy \Id.  During procedure {\sc
Insert}, only a constant number of pebbles are {\em touched} -- i.e.  added or
reassigned. Let $\Gamma$ be the situation before {\sc Insert} and $\Gamma'$ be
the situation after {\sc Insert} finished. Let $P$ be a free place in $\Gamma'$
and $X$ be the smallest pebble to the right of $P$. We distinguish two cases:

\vskip 1mm
\noindent
{\bf Case 1: $X$ was touched during {\sc Insert}}\\[.1mm]
If $p<x/2$, no pebbles are required on $P$, so let us suppose that $p\ge x/2$.
However, since $\Gamma'$ is valid, the free bandwidth before $X$ is less than
$x$, so $p=x/2$, and there is only one pebble required in $P$; this pebble will
be placed on $P$ and charged to $X$. Obviously, for each touched pebble $X$,
there may be only one free place of size $x/2$ to the left (because of \Ia), so
every touched pebble will be charged at most one coin using in total constant
number of coins.

\vskip 1mm
\noindent
{\bf Case 2: $X$ was not touched during {\sc Insert}}\\[.1mm]
If $P$ was free in $\Gamma$ the required amount of $\left\lfloor
2p/x\right\rfloor$ coins was already present on $P$ in $\Gamma$, so let us
suppose that $P$ was not free in $\Gamma$. That means that $P$ became free in
the course of {\sc Insert} when some pebbles were reallocated. Using similar
arguments as in the previous case we argue that $p=x/2$. However, during {\sc
Insert} only a constant number of free places of a given size could be created,
so it is affordable to put one coin on each of them.
\end{proof}

\begin{lemma}
\label{lemma:deleteComplexity}
Let us suppose that procedure {\sc Delete} was called from a situation in which
invariants \Ia -- \Id hold.  Then it is possible to add a constant number of
coins, remove one coin per iteration of the main loop, and reallocate the
remaining coins in such a way that invariants \Ia -- \Id remain valid. 
\end{lemma}

\begin{proofsk}
Let us suppose that there is at least one full iteration of
the main loop.  Recall the notation from the proof of
Theorem~\ref{theorem:deleteCorrect}, i.e. we number the iterations of the main
loop, and $\Gamma_t$ is the configuration at the beginning of $t$-th iteration.
Moreover, $\Gamma_t'$ is obtained from $\Gamma_t$ by putting a new pebble $A_t$
on $P_t$. From the proof of the theorem it follows that $\Gamma'_t$ is always a
valid situation. We prove by induction on $t$ that the following can be
maintained:

\vskip 2mm
\noindent
{\em
For every iteration $t>1$ of the {\bf while} loop, the corresponding
$\Gamma'_t$ satisfies \Ia--\Id, all pebbles to the right of $i_t$ have size at least
$a_t$, and one extra coin lays on $A_t$. Moreover, if some
pebble to the right of $i_t$ has the size $a_t$, then two extra coins lay on
$A_t$.
}

We omit the details about the induction basis, and proceed with the induction step.
In order to prove the claim for $\Gamma_{t+1}'$, consider
the situation when the algorithm finishes the $t$-th iteration and enters the
$t+1$st.  $\Gamma_{t+1}'$ is obtained from $\Gamma_t'$ by removing $A_t$,
moving $X_t$ to $i_t$, and placing $A_{t+1}$ on $i_{t+1}=j_t$. Note that in
this case $x_t\ge a_t$, so there is no swap. It is possible to show
that all pebbles to the right of $i_{t+1}$ have size at least $a_{t+1}$, and
that there is no free place between $i_t$ and $j_t$ in $\Gamma_{t+1}'$. Since
for the free places before $i_t$ and after $j_t$, \Id remains valid, we argue
that \Id holds in $\Gamma_{t+1}'$.

Now we show how to find two free coins -- one to pay for the current
iteration, and one to be placed on $A_{t+1}$. If $x_t=a_t$, there are two
coins placed on $A_t$. Otherwise (i.e. in case that $x_t>a_t$) one coin comes from the deletion
of $A_t$ and the other can be found as follows. Since $X_t$ was placed on $i_t$,
and $x_t>a_t$, there must have been a free place $P$ of size $x_t/2$ in
$\Gamma_t'$. Moreover, $X_t$ was to the right of $P$, and so there must have
been at least one coin on $P$. In $\Gamma_{t+1}'$, $P$ is covered by $X_t$, so
the coin can be used.

The last thing to show is to find a second free coin in case that there
exists some pebble $Q$ to the right of $A_{t+1}$ of size $q=a_{t+1}$. In
this case $X_t$ is white in $\Gamma_t'$ -- otherwise it would hold that
$a_{t+1}=x_t$ and there would be no pebble of size $x_t$ to the right of $X_t$.
Hence $x_t\le a_{t+1}/2$ and $A_{t+1}$ in $\Gamma_{t+1}'$ covers a place of
size $a_{t+1}/2$ that has been free in $\Gamma_t'$. According to \Id 
there is a coin on this place in $\Gamma_t'$; this coin can be used.

The proof of the theorem is concluded by considering the last iteration. Let
$\Gamma_{t_{fin}}'$ be the last situation. The final situation is obtained from
$\Gamma_{t_{fin}}'$ by removing $A_{t_{fin}}$.  We show that \Id holds. The
only free place that could violate \Id is the one remained after $A_{t_{fin}}$,
however, there was a coin on $A_{t_{fin}}$, and all pebbles to the right (if
any) are bigger than $a_{t_{fin}}$.
\end{proofsk}

\section{Conclusion}

We have presented an online algorithm for bandwidth allocation in wireless networks,
which can be used to perform the OVSF code reallocation with the amortized complexity
of $O(1)$ reallocations per request. This is an improvement over the previous
best known result achieving the competitive ratio of $O(n)$. Moreover, the constant
in our algorithm is small enough to be of practical relevance.

On the other hand, no attempt has been made at minimizing this constant. With the best
known lower bound of $1.5$ it would be worthwhile to close the gap even further.

\bibliography{references}

\begin{thebibliography}{1}

\bibitem{ASO97}
F.~Adachi, M.~Sawahashi, and K.~Okawa.
\newblock Tree structured generation of orthogonal spreading codes with
  different lengths for the forward link of {DS-CDMA} mobile radio.
\newblock {\em IEE Electronic Letters}, 33(1):27--28, 1997.

\bibitem{BDM05}
G.~S. Brodal, E.~D. Demaine, and J.~I. Munro.
\newblock Fast allocation and deallocation with an improved buddy system.
\newblock {\em Acta Informatica}, 41(4--5):273--291, March 2005.

\bibitem{DCC05}
D.~C. Defoe, S.~R. Cholleti, and R.~K. Cytron.
\newblock Upper bound for defragmenting buddy heaps.
\newblock In {\em LCTES '05: Proceedings of the 2005 ACM SIGPLAN/SIGBED
  conference on Languages, compilers, and tools for embedded systems}, pages
  222--229, New York, NY, USA, 2005. ACM Press.

\bibitem{E+04}
T.~Erlebach, R.~Jacob, M.~Mihal{\'a}k, M.~Nunkesser, G.~Szab{\'o}, and
  P.~Widmayer.
\newblock An algorithmic view on {OVSF} code assignment.
\newblock In V.~Diekert and M.~Habib, editors, {\em STACS}, volume 2996 of {\em
  Lecture Notes in Computer Science}, pages 270--281. Springer, 2004.

\bibitem{K65}
K.~C. Knowlton.
\newblock A fast storage allocator.
\newblock {\em Communications of ACM}, 8(10):623--624, 1965.

\bibitem{MS00}
T.~Minn and K.-Y. Siu.
\newblock Dynamic assignment of orthogonal variable-spreading-factor codes in
  {W-CDMA}.
\newblock {\em IEEE Journal on Selected Areas in Communications},
  18(8):1429--1440, 2000.

\bibitem{W}
{Wikipedia}.
\newblock Universal mobile telecommunications system.
\newblock Available at: \url{http://en.wikipedia.org/wiki/Umts}.
\newblock From Wikipedia, the free encyclopedia [Online; accessed 23. March
  2006].

\end{thebibliography}

\newpage
\section*{Appendix}

This part contains the technical parts and proofs excluded from the
previous sections. Let us first present a few structural lemmata describing valid situations. Informally,
we prove that a valid situation (i.e. situation satisfying invariants \Ia--\Ic) can be described as follows:
assign black color to pebbles that are not preceded by a bigger pebble. Then these pebbles form a
non-decreasing sequence with as few free places between pebbles as possible. In each free place of this sequence, at
most one white pebble may be present, aligned to the left. Moreover, white pebbles form a strictly increasing sequence.

Unless stated otherwise, the following lemmas assume a valid situation.

\begin{obs}\label{obs:blackIncreasing}
The sizes of black pebbles form a non-decreasing subsequence.
\end{obs}

\begin{obs}\label{obs:removeLast}
If the last pebble is removed, the invariants \Ia--\Ic remain true, i.e. the
situation remains valid.
\end{obs}

\begin{lemma}\label{lemma:empty}
In a non-empty valid situation, the first pebble is placed at position 0.
\end{lemma}
\begin{proof}
Consider, for the sake of contradiction, the first pebble $A$ of size $a$. 
Clearly, $A$ must be at position $ia$ for some $i>0$. But then there is some
free place of size $a$ before $A$ -- contradiction with \Ia.
\end{proof}

\begin{lemma}\label{lemma:sequenceBlack}
All black pebbles of a given level $l$ occupy consequent places.
\end{lemma}
\begin{proof}
Let $A$ and $B$ be two black pebbles of level $l$ such that $a=b=2^l$, $A$ being
on the left.  Let $X$ be a pebble between $A$ and $B$. Because $B$ is black,
$x\le b$ holds. From \Ic it follows that $X$ must be black, i.e. $x\ge a$.
Hence, $X$ is black and of the same size as $A$ and $B$. That means that the
black pebbles of level $l$ form a sequence that can be interrupted by some free
places but not by pebbles of different sizes.

Consider now, for the sake of contradiction two black pebbles $A$ and $B$
of level $l$ separated by some free places. Because the positions and sizes of $A$ and $B$
are multiples of $2^l$, it follows that if they are separated by some free places, 
there is also a free place of level $l$ between them. However, this would contradict
\Ia.
\end{proof}

\begin{lemma}\label{lemma:spaceWhite}
The left neighbor of a white pebble $A$ always exists, is black, and strictly bigger than $A$. 
Moreover, there is no free space between $A$ and its left neighbor.
\end{lemma}
\begin{proof}
Consider a white pebble $A$. Since there is a bigger pebble before $A$, there must be
a left neighbor $B$. Because of \Ib, $B$ is black. If $b\le a$ then $A$ would be black, too,
so $B$ must be bigger than $A$. 

Let $A$ and $B$ be separated by some free places. Since $b$ is a multiple of $a$, 
there must be a free place of size $a$ between them -- a contradiction with \Ia.
\end{proof}

\begin{corollary}\label{lemma:spaceRight}
If there is a free place before a pebble $A$, then $A$ is black.
\end{corollary}

\begin{lemma}\label{lemma:spaceSize}
Consider a pebble $A$ followed by a free place. Then the overall size of these free places is $p\ge a$.
\end{lemma}
\begin{proof}
Let $A$ be a pebble of level $l$ followed by some free place $P$.
If $A$ is the last pebble, then $A$ is followed by a place of level $l$ which
is free.

Let $B$ be the right neighbor of $A$. Because of
Corollary~\ref{lemma:spaceRight}, $B$ must be black, i.e. $b\ge a$. Since the
positions and sizes of both $A$ and $B$ are multiples of $a=2^l$, the free
place following $A$ has to be at least of size $a$.
\end{proof}

\begin{lemma}\label{lemma:whiteSpaceAfter}
Consider a black pebble $B$ followed by a white pebble $A$. Then $A$ is followed by
free places of overall size at least $b-a\ge b/2\ge a$.
\end{lemma}
\begin{proof}
Because of Lemma~\ref{lemma:spaceWhite}, there is no free space between $A$ and $B$, 
and $b\ge2a$. If $A$ is the last pebble, the proposition follows immediately.

\myfig{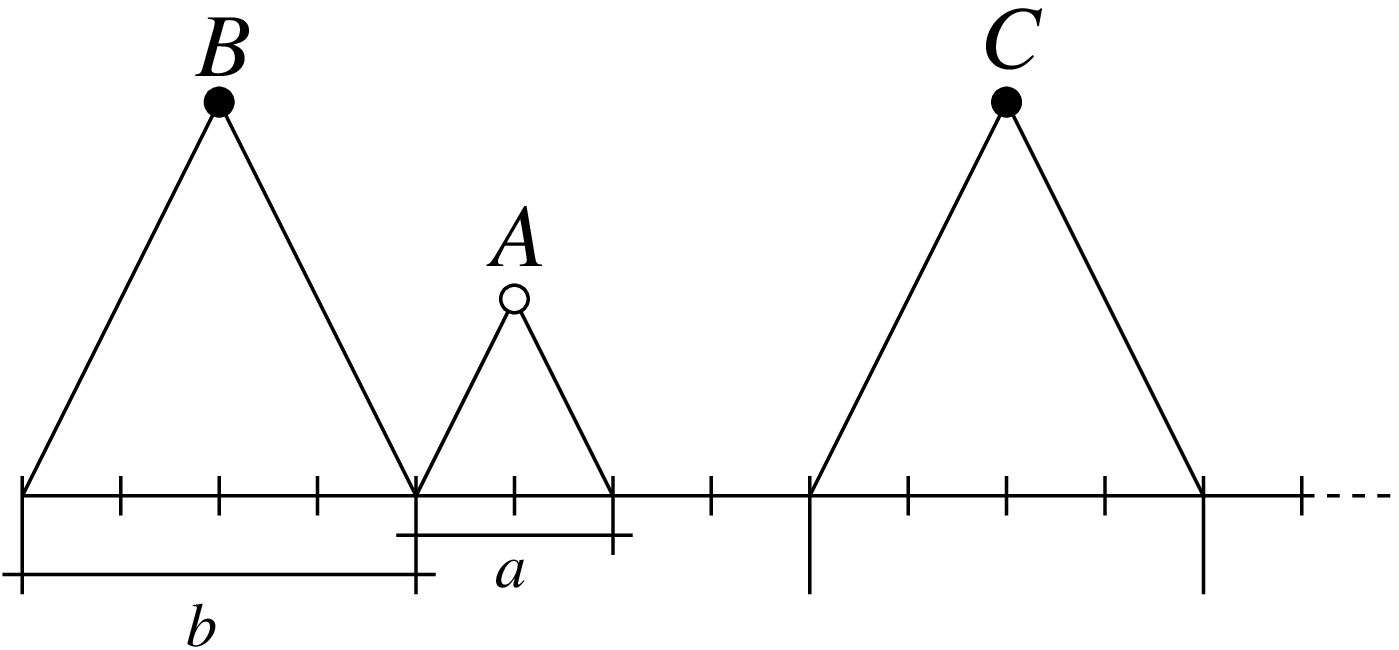}{6cm}{Situation in the proof of Lemma~\ref{lemma:whiteSpaceAfter}.}

If $A$ is not the last pebble then there exists its right neighbor $C$ (this
situation is depicted in Figure~\ref{fig:004}). Because
of \Ib, $C$ is black, i.e. $c\ge b$. As the positions and sizes of $B$ and $C$
are multiples of $b$, the distance between them is at least $b$. Because
$A$ immediately follows $B$, and there are no other pebbles between $B$ and $C$,
the overall size of free spaces following $A$ is at least $b-a$, and the
proposition follows.
\end{proof}

\begin{lemma}\label{lemma:whiteIncreasing}
The sizes of white pebbles form a strictly increasing subsequence.
\end{lemma}
\begin{proof}
Consider a white pebble $A$, and a white pebble $B$ located somewhere after $A$.
Consider, for the sake of contradiction, that $a\ge b$. Because of
Lemma~\ref{lemma:whiteSpaceAfter}, the place $P$ of size $a$ immediately
following $A$ is free. Obviously, $B$ is located after $P$ which is a
contradiction with \Ia.
\end{proof}

The following two lemmas follow directly from the structure of the valid situations,
and provide a useful tool in proving the correctness of the algorithm:

\begin{lemma}\label{lemma:whiteKiller}
Consider a black pebble $B$ immediately followed by a white pebble $A$. Then there
is no white pebble $X\not=A$ such that $a\le x<b$.
\end{lemma}
\begin{proof}
Consider, for the sake of contradiction, a white pebble $X$ satisfying $a\le x<b$.
Because of Lemma~\ref{lemma:whiteIncreasing}, $X$ cannot be located before $A$.
However, according to Lemma~\ref{lemma:whiteSpaceAfter}, there are free spaces of
overall size at least $b/2$ immediately following $A$. Because $x<b$, it holds $x\le b/2$,
and \Ia is violated.
\end{proof}

\begin{lemma}\label{lemma:insert}
For a given $l$, let $P$ be the first free place of size $p=2^l$. Then the free bandwidth
before $P$ is strictly less than $p$.
\end{lemma}
\begin{proof}
First consider the case when all pebbles to the left of $P$ are of size at least
$p$. Since all the positions and sizes of those pebbles are multiples of $p$,
there is either no free place before $P$, or a free place of level at least
$l$. However, the latter case would contradict the fact that $P$ is the first
free place of size $p$.

\myfig{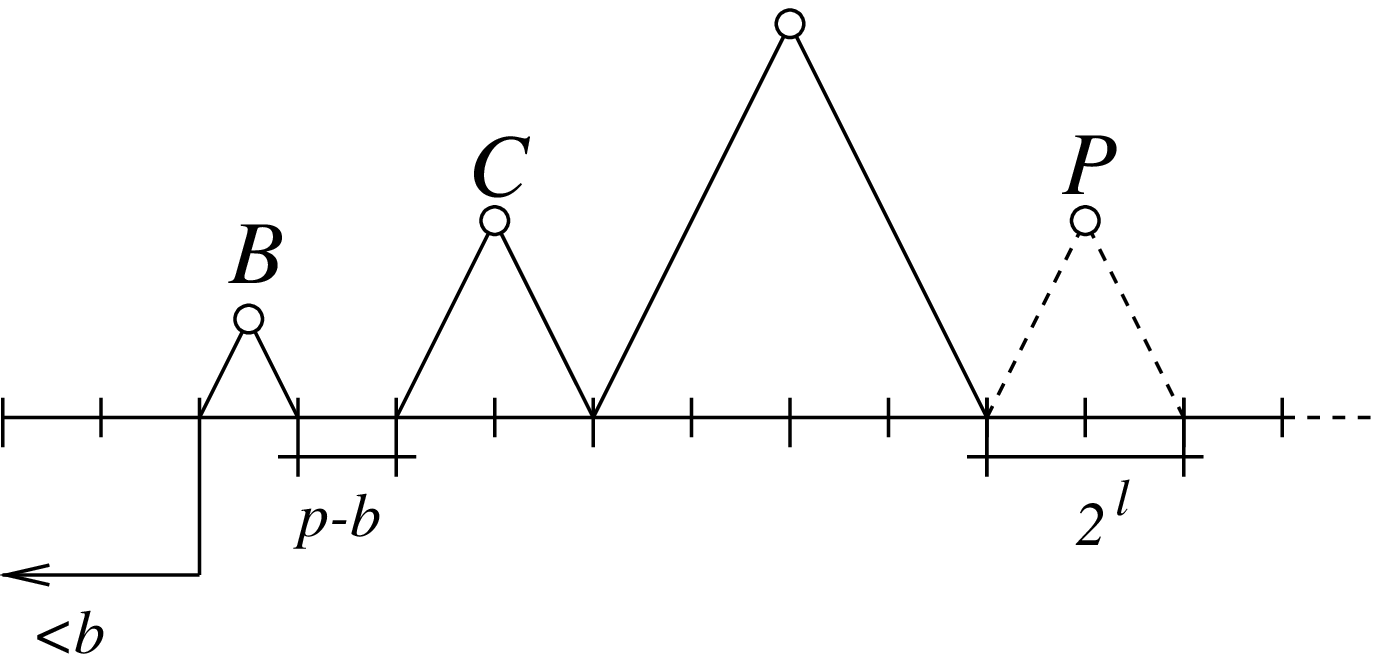}{6.5cm}{Situation in the proof of Lemma~\ref{lemma:insert}: there is a smaller pebble $B$ to the left of $P$.}

Let $B$ be the first pebble to the left of $P$ such that $b<p$, i.e. there are
only pebbles of size at least $p$ between $B$ and $P$ (see Figure~\ref{fig:005}). From \Ia it follows that
the free bandwidth before $B$ is strictly less than $b\le p/2$.
If there are any pebbles between $B$ and $P$, let $C$ be $B$'s right neighbor.
Using the same argument as above we can argue that there is no free place
between $C$ and $P$. What remains to be shown is that the overall size of free
places between $B$ and $P$ immediately following $B$ is at most $p-b$. Denote
this size as $\alpha$ and let us suppose, for the sake of contradiction, that
$\alpha>p-b$. Since the positions and sizes of $P$ and all pebbles between $B$
and $P$ are multiples of $p$, it follows that $\alpha\ge p$. The reason why
it is so is clear from Figure~\ref{fig:006}.
\myfig{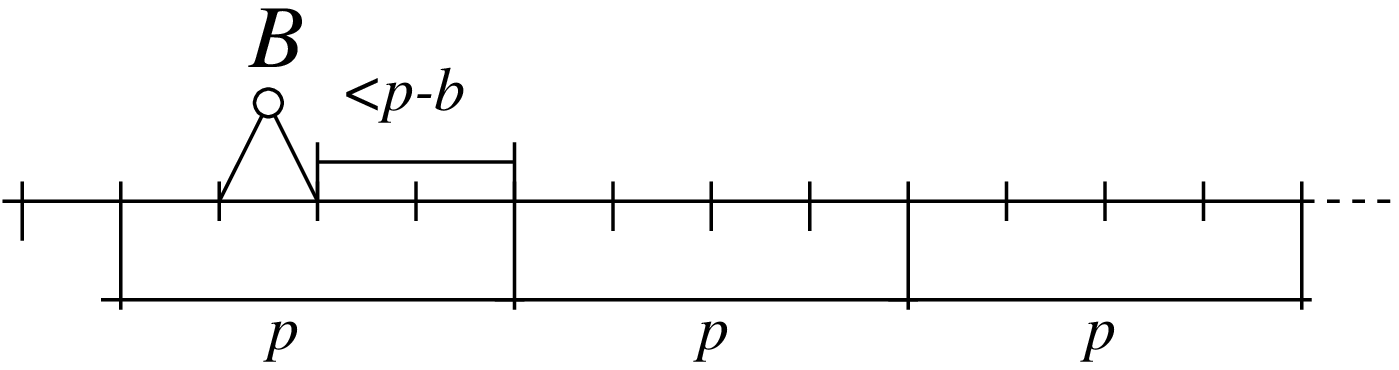}{6cm}{Situation in the proof of Lemma~\ref{lemma:insert}: if $\alpha>p-b$, there is a free place of size $p$.}
Consider the places of size $p$: $B$ must be fully located in one of them, and
the remaining free places within this place have overall size at most $p-b$.
Since $\alpha>p-b$ there must be some free place in the next place of size $p$. However,
since all subsequent pebbles are of size at least $p$, they start at the
beginning of a place of size $p$. Hence the next place of size $p$ is free
-- a contradiction with the fact that $P$ is the first free place of size $p$.
\end{proof}

With the developed machinery we are able to prove the crucial Lemma~\ref{lemma:canPut},
stating that in valid situations, if there is enough free bandwidth to satisfy
an insertion request of size $a$, there is always a free place of size $a$.

\begin{proofof}{Lemma~\ref{lemma:canPut}}
Consider, for the sake of contradiction, that there is no free place of size
$a$. Let $P$ be the last place of size $a$, i.e. the interval $\lla
2^n-a,2^n-1\rra$. Obviously, $P$ is not free. If there is a pebble $B$ of size
$b<a$ inside $P$ then because of \Ia, the free bandwidth before $B$ is strictly
less than $b$.  However, the free bandwidth after $B$ is at most $a-b$, hence
there is strictly less than $a$ free bandwidth overall.

So it must be that $P$ is not free because it is a part of a pebble $X$ of size
at least $a$. Following Observation~\ref{obs:removeLast}, $X$ can be removed
and the invariants still hold. Now consider the first free place $A$ of size
$a$ (it is at the beginning of the removed pebble $X$): because of
Lemma~\ref{lemma:insert}, there is strictly less than $a$ free bandwidth before
$A$. However, this is all free bandwidth that exists in the original situation.
\end{proofof}

Another observation that is useful in the case-analysis of the proof of Theorem~\ref{thm:insertCorrect}
is the following lemma:

\begin{lemma}\label{lemma:rightNeighbor}
Consider a sequence of pebbles and free places that forms a valid situation.
Let $P$ be the first free place of size $p$. Let $Z$ be the right
neighbor of $P$.  If $Z$ exists, it is black and $z>p$.
\end{lemma}
\begin{proof}
Since $P$ is a free place and $Z$ is $P$'s right neighbor, there is a free
place immediately preceding $Z$.  According to Lemma~\ref{lemma:spaceWhite} $Z$
cannot be white.
The fact that $z>p$ follows immediately from \Ia.
\end{proof}

Now we are ready to prove the correctness of procedure {\sc Insert}:

\begin{proofof}{Theorem~\ref{thm:insertCorrect}}
Consider an insertion request of size $a$, and suppose the situation is valid.
Because of Lemma~\ref{lemma:canPut} the line \ref{alg:insert.1} of Algorithm
\ref{alg:insert} is correct, i.e.  there is a free place of size $a$. Let $P$
be the first such free place.

If $P$ does not have a left neighbor, the algorithm puts $A$ on $P$ in line
\ref{alg:insert.2} and exits. Since there is no pebble to the left of $P$ and
$P$ is the first free place if size $a$, it means that $P$ starts at position
$0$. However, due to Lemma~\ref{lemma:empty} it follows that there are no other
pebbles, and the situation is valid.
From now on suppose that $P$ has a left neighbor $B$, and denote the potential
right neighbor of $P$ by $Z$.  Here we present the full case analysis.

\vskip 3mm
\noindent
{\bf Case 1: $B$ is black}\\[1mm]
The algorithm puts $A$ on $P$ in line \ref{alg:insert.2} and exits.  Obviously,
no pebble to the left of $A$ changes color.  If there are some pebbles to the
right of $A$, then because of Lemma~\ref{lemma:rightNeighbor} there exists a
black $Z$, such that $z>a$ (see Figure~\ref{fig:007}). Hence, no pebble changes
color after $A$ is put on $P$.  We prove that all invariants hold:
\begin{itemize} 
\item[\Ia:] For $A$ it follows from Lemma~\ref{lemma:insert}. For all other
pebbles the free bandwidth before them could have only been decreased.
\item[\Ib:] Since $B$ is black, the only way \Ib could be violated is if $Z$
exists and is white.  However, due to Lemma~\ref{lemma:rightNeighbor} it is not
possible.
\item[\Ic:] Consider, for the sake of contradiction, that after $A$ was put on
$P$, the invariant \Ic is violated. Assume that $A$ is white. In this case $Z$
exists, is black and $z=b$, which contradicts  Lemma~\ref{lemma:rightNeighbor}.
On the other side if $A$ is black, then $B$ or $Z$ must be white. This
contradicts Lemma~\ref{lemma:rightNeighbor}, too.
\end{itemize}

\myfig{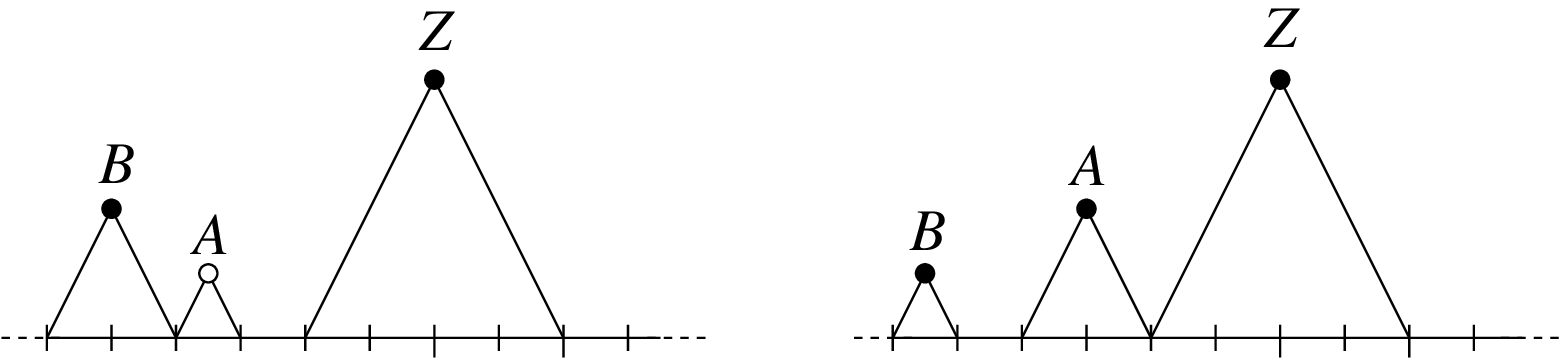}{12cm}{Two examples of Case 1}

\noindent
{\bf Case 2: $B$ is white}\\[1mm]
Because of Lemma~\ref{lemma:spaceWhite}, there exists a left neighbor $C$ of
$B$, such that $C$ is black and $c>b$. Hence, line~\ref{alg:insert.3} is
correct. We again distinguish three cases according to the relation of $c$ and
$a$.

\vskip 3mm
\noindent
{\em Case 2.1: $c<a$}\\[1mm]
This case is handled by putting $A$ on $P$ in line~\ref{alg:insert.4}. Because
$C$ is black and $b<c<a$, $A$ is colored black. Obviously, no pebble before $A$
changed color. If there are some pebbles to the right of $A$, then because of
Lemma~\ref{lemma:rightNeighbor} there exists a black $Z$, such that $z>a$ (see
Figure~\ref{fig:008}).  Hence, no pebble changes color after $A$ is put on $P$. 

\myfig{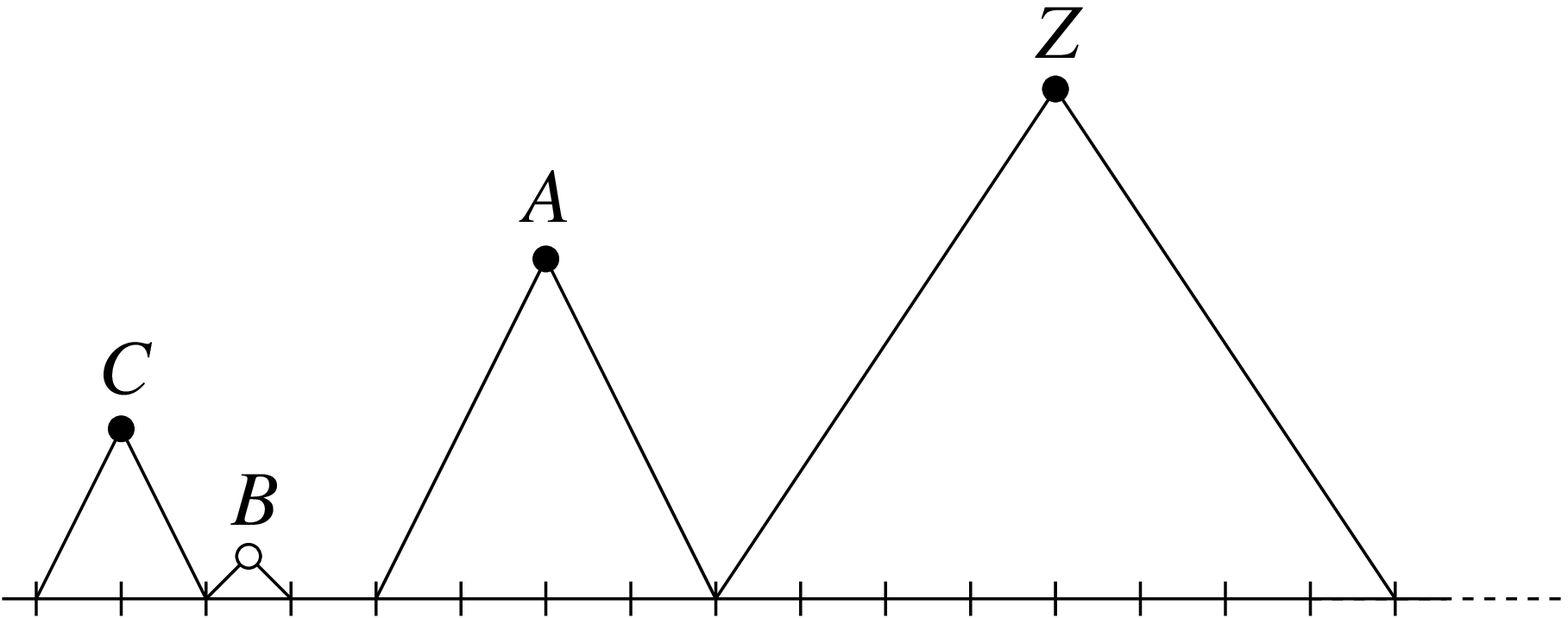}{7.5cm}{Case 2.1}

We prove that the situation remains valid, i.e. the invariants \Ia--\Ic remain true.
\begin{itemize}
\item[\Ia:] The same argument as in {\bf Case 1}.
\item[\Ib:] Holds trivially, since only a black pebble $A$ has been added.
\item[\Ic:] Consider, for the sake of contradiction, that after $A$ was put on
$P$, there is a white pebble $W$ between two black pebbles of the same size.
Since $A$ is black, $A\not=W$, i.e. $W$ is a neighbor of $A$. Since $Z$ is
black, we have $W=B$. As $c<a$, invariant $\Ic$ holds -- a contradiction.  
\end{itemize}

\vskip 3mm
\noindent
{\em Case 2.2: $c=a$}\\[1mm]
This situation is handled in the block on lines
\ref{alg:insert.5}--\ref{alg:insert.6}: first, pebble $B$ is temporarily
removed. Since there has been no other pebble between $A$ and $C$, and $a=c$,
the place of size $a$ immediately following $C$ is now free.  Put $A$
immediately after $C$. Since $b<c=a$, the place of size $b$ immediately
following $A$ is now free, so $B$ can be put there (see Figure~\ref{fig:009}).
Since $C$ is black, $A$ is black, too.  Using arguments as in Case 2.1 we argue
that no pebble has changed its color.

\myfig{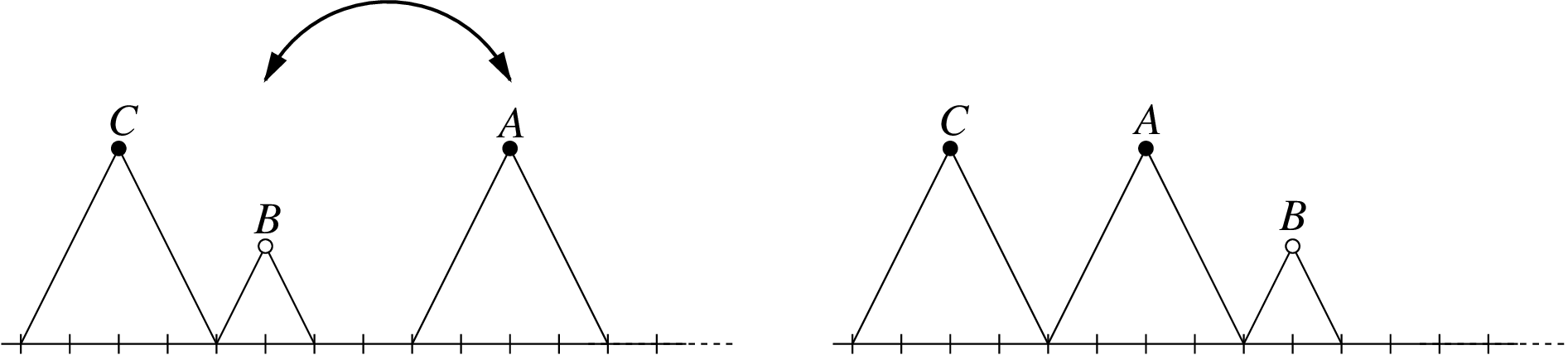}{13cm}{Case 2.2}

Again, we prove that after the reassignment all invariants hold.
\begin{itemize}
\item[\Ia:] Suppose that $A$ was put on $P$. The invariant holds using the same
argument as in Case 1.  The reassignment changed the status of pebbles $A$ and
$B$ only. However, the free bandwidth before $B$ has not changed, and the free
bandwidth before $A$ could have only been decreased.
\item[\Ib:] Suppose for the sake of contradiction that after the reassignment
of pebbles there are two neighboring white pebbles. Clearly, one of them must be
$B$.  However, $A$ is black so it must be that $B$'s right neighbor is white.
The contradiction follows from Lemma~\ref{lemma:rightNeighbor} -- $B$'s right
neighbor is the original $P$'s right neighbor $Z$, which is black.
\item[\Ic:] Suppose that after the algorithm finishes, there is a white pebble
$W\not= A$ between two black pebbles of the same size. The only candidate for
$W$ is the pebble $B$. Due to Lemma~\ref{lemma:rightNeighbor}, if the right
neighbor of $B$ exists, $z>a$ holds, which is a contradiction.
\end{itemize}

\vskip 3mm
\noindent
{\em Case 2.3: $c>a$}\\[1mm]
The algorithm starts this case by temporarily removing $B$.  First, we argue
that we can without loss of generality consider $a\ge b$.  Suppose that $a<b$,
i.e. the situation is as on Figure~\ref{fig:011} left. We will treat this
situation exactly as if $a$ and $b$ would be swapped, i.e. as if the requested
size was $b$ and the existing pebble $B'$ was of size $a$ (Figure~\ref{fig:011}
right) -- the removal of $B$ and $B'$ will render the same situation.

\myfig{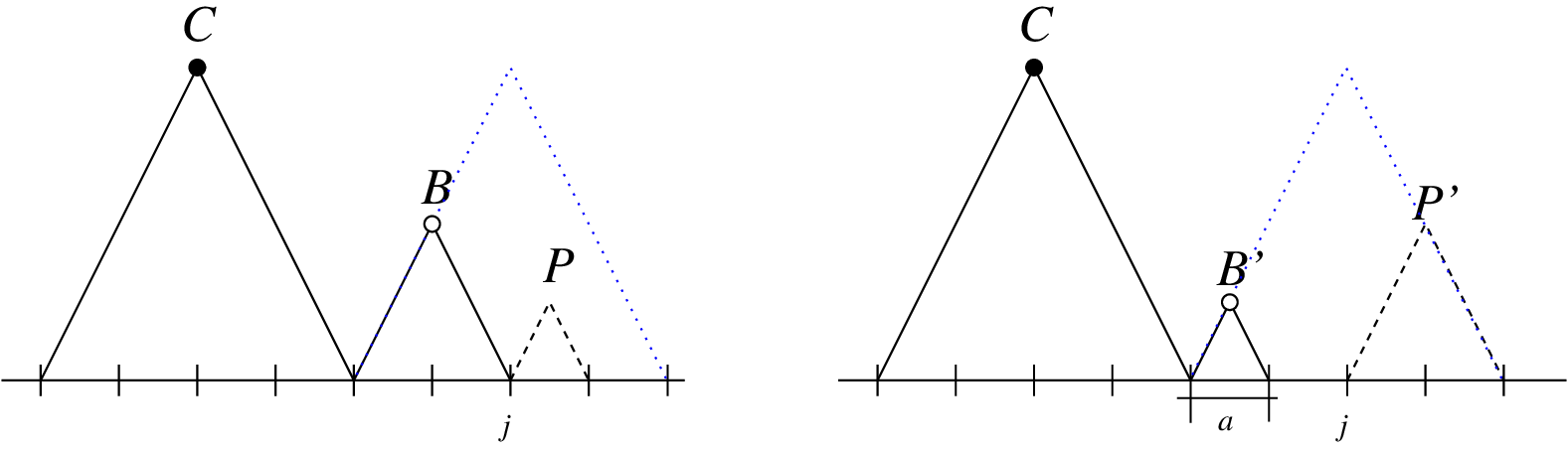}{11cm}{Case 2.3: the situation $a<b$ can be transformed to $a\ge b$.}

However, we have to show that if we replace the pebble $B$ of size $b$ by a
pebble $B'$ of size $a$, the situation remains valid (i.e.  \Ia--\Ic still
hold). Clearly, the colors of all pebbles remain the same, and hence \Ib, \Ic
remain true.  \Ia remains true because Lemma~\ref{lemma:insert} states that the
free bandwidth before $P$ (and hence before $C$) is less than $a$. Let $j$ be
the position where $P$ starts.  What remains to be shown is the fact that the
first free place of size $b$ starts at $j$. Since $c>b>a$, and there is no free
space between $C$ and $B$ and between $B$ and $P$, both $B$ and $P$ fit into a
place of size $c$ immediately following $C$.  However, due to
Lemma~\ref{lemma:whiteSpaceAfter}, the mentioned space of size $c$ does not
contain other pebbles than $B$.  Since $b\le c/2$, there is a free place $P'$ of
size $b$ starting at $j$.  $P'$ is the first free place of size $b$: because of
Lemma~\ref{lemma:insert}, the free bandwidth before  $P$ (and hence before $C$)
is less than $a$; the free bandwidth between $B'$ and $P'$ is $b-a$, hence no
free space of size $b$ exists before $P'$.

From now on let us suppose $a\ge b$. Let $i$ be the closing position of size
$a$. We argue that there must be a pebble at $i$.  To see why recall the
definition of closing position: either there is a pebble of size $a$ ending at
$i-1$, or a pebble of size $>a$ starting at $i$. Hence, if there is no pebble at
position $i$, there is a pebble of size $a$ just before $i$, and due to
Lemma~\ref{lemma:spaceSize} there is a free space of size $a$ immediately
following it. However, $i$ is to the left of $C$ -- a contradiction with the
fact that $P$ was chosen to be the first free place of size $a$.
Let us denote the pebble at position $i$ by $D$, and once more distinguish two
sub-cases:

\vskip 3mm
\noindent
{\sf Case 2.3.1: $D$ is white}\\[1mm]
From the definition of the closing position it follows that $D$ is a white
pebble immediately preceded by a black pebble $X$ of size $a$. Obviously, $D$ has
a right neighbor $E$ which is, due to \Ib black\footnote{as a special case, it
might be $E=C$}, and due to \Ic it holds $e>a$. Moreover, it holds that $a=b$:
we argued above that $a\ge b$. Because of Lemma~\ref{lemma:whiteKiller} applied
to $X$ and $D$, and Lemma~\ref{lemma:whiteIncreasing}, it holds that $b\ge a$.


The action of the algorithm in this case is depicted on Figure~\ref{fig:010}:
$E$ is moved after $C$, and $A$, $B$ are inserted between $X$ and $D$. We first
prove that this operation is correct, i.e. there is always an appropriate free
place to put the pebbles. 

After $D$ and $E$ were removed, the three places of size $a$ immediately
following $X$ are free; the reason is that $e\ge 2a$, i.e. $E$'s starting
position was a multiple of $a$: the first of the three places originally
contained only $D$, and the remaining two were occupied by $E$.  Hence, $A$,
$B$, and $D$ can be placed after $X$. Moreover, after $B$ was removed, the free
place of size $e$ immediately following $C$ is free -- if $Z$ exists, it is due
to Lemma~\ref{lemma:rightNeighbor} black, and so its starting position is a
multiple of $c$. Since $e\le c$, it follows that $E$ can be placed at $B$'s
original position.
We now make sure that the situation is valid, i.e. the invariants \Ia--\Ic
hold. First, let us argue about
the colors of the pebbles: pebbles to the left of $X$ (including $X$) don't
change color, and neither do pebbles to the right of $Z$ (including $Z$).
Because $a=b$, pebbles $A$ and $B$ become black. Pebble $D$ remains white. Since
$c>a$, $C$ remains black. $E$ may be either white, if $e<c$, or black, if
$e=c$. 

Lemma~\ref{lemma:whiteKiller} applied to $B$ and $C$ ensures that there are no
white pebbles of size between $a$ and $c$. Since $a<e\le c$, all pebbles between
$E$'s original position and $C$ were black. Hence, they remain black also after
the reassignment.  Now, let us argue about the invariants:
\begin{itemize}
\item[\Ia:] For pebble $X$ and all pebbles to the left \Ia holds trivially. For
$Z$ and pebbles to the right the free bandwidth before them decreased when $A$
was added. For $A$ and $B$, \Ia holds because Lemma~\ref{lemma:insert} asserts
that the free bandwidth before $P$ is less than $a$, and they were placed to
the left of $P$. For $D$, the free bandwidth to the left did not change.

Consider the new position of pebble $E$. Since $E$ was placed at $B$'s original
position, the bandwidth before $E$ is the original free bandwidth before $B$
(due to Lemma~\ref{lemma:insert} at most $a$) increased by $e$ ($E$ was
removed), and decreased by $2a$ ($A$ and $B$ were inserted). Hence, we get that
the free bandwidth before $E$ is at most $a+e-2a<e$, and \Ia holds for $E$.

As we argued before, all pebbles between $E$'s original position and $E$'s new
position are black, and bigger or equal than $E$, thus \Ia holds for them, too.
\item[\Ib:] Violating \Ib means that there are two consecutive white pebbles.
Obviously, if such two pebbles exist they must be between $X$ and $Z$ because
the original position was valid. However from these pebbles, only $D$ and $E$
may be white, and only if $e<c$. But in this case there is a black pebble
$C\not=E$ between $E$ and $D$.
\item[\Ic:] Again, only $D$ and $E$ are candidates for violating \Ic.  If $Z$
exists, $z>c$, because originally there was white $B$ between them. Hence $E$
cannot violate \Ic.
Originally, $X$ was the last black pebble of size $a$, so $D$ cannot violate
\Ic, either.
\end{itemize}

\vskip 3mm\goodbreak
\noindent
{\sf Case 2.3.2: $D$ is black}\\[1mm]
In this case the definition of closing position ensures that $D$ is the first
black pebble of size bigger than $a$, and that there are no black pebbles of size
$a$. Thus  we get the following inequalities $$b\le a<d\le c$$ As a consequence
of Lemma~\ref{lemma:whiteKiller} applied to $B$ and $C$ we get that all pebbles
between $D$ and $C$ are black (with a possible special case $D=C$).


The action of the algorithm is described on Figure~\ref{fig:012}: pebble $D$ is
removed, and pebbles $A$ and $B$ are placed into the free space.  Pebble $D$ is
then put at $B$'s original position. First, we reason that this action is well
defined. Since both $a,b\le d/2$, pebbles $A$ and $B$ fit into the free space
created by removing $D$. Moreover, the place of size $c$ immediately following
$C$ is, after removing $B$, free: if $Z$ exists, it starts at a position that is
a multiple of $c$. Since $d\le c$, $D$ fits into this position.

Now consider the colors of the pebbles: pebble $A$ becomes black, since $D$ was
the first (black) pebble of size bigger than $a$. $B$ becomes either black or
white, and so does $D$. The colors of all other pebbles do not change. Again,
let us argue about the invariants:

\begin{itemize}
\item[\Ia:] For $A$, \Ia holds because, due to Lemma~\ref{lemma:insert} the
free bandwidth before $P$ was at most $a$. For pebbles to the left of $A$
nothing changed. For $Z$ and all pebbles to the right the free bandwidth
decreased, and so it did for $B$. 

Consider the free bandwidth before $D$: because of \Ia, the free bandwidth
before the original position of $B$ was less than $b$. Hence the free bandwidth
before $D$ is less than $b+d-a-b<d$. All pebbles between $D$'s original position
and $D$'s new position are black and bigger than $D$ so \Ia holds for them, too.
\item[\Ib:]The only possibility to violate \Ib is that $B$ and $D$ are
consecutive white pebbles. However, $B$ becomes white only if there is a black
$C\not=D$ before it, because $d>a$.
\item[\Ic:]Since $z>c$, \Ic cannot be violated by $D$. Moreover, $a<d$ so \Ic
cannot be violated by $B$.
\end{itemize}
\end{proofof}

\noindent
The following lemma is used in the proof of correctness of {\sc Delete}:

\begin{lemma}\label{lemma:whiteKiller2}
Consider a valid situation with a black pebble $B$ starting at location $i$,
immediately followed by a white pebble $A$. Let $j$ be the closing position of
size $2a$. Then no free place and no white pebble starts at any location between
$j$ and $i$.
\end{lemma}
\begin{proof}
First let us prove that there is no free place between $j$ and $i$.  Consider,
for the sake of contradiction the leftmost free place starting between $j$ and
$i$.  Such a free place is not unique, so let $P$ be the one with the biggest
level among them.  If $P$ starts at $j$, then, by the definition of closing
position, it is immediately preceded by a black pebble of size $2a$. It follows
from Lemma~\ref{lemma:spaceSize} that $P$ is of size at least $2a$ -- a
contradiction with \Ia applied to $A$. Hence, $P$ must start to the right of
$j$, and is immediately preceded by a pebble $X$ (because it is leftmost). If
$X$ is black, then $x\ge2a$, and the same contradiction as above follows. If
$X$ is white, due to Lemma~\ref{lemma:whiteIncreasing}, $x<a$. Moreover, due to
Lemma~\ref{lemma:spaceWhite}, $X$ is immediately preceded by a black pebble
$Y$, such that $y\ge 2a$. Finally, due to Lemma~\ref{lemma:whiteSpaceAfter}
there is at least $y-x>a$ free bandwidth immediately following $X$ -- a
contradiction.

Finally, we argue that there is no white pebble between $j$ and $i$. Due to
Lemma~\ref{lemma:whiteSpaceAfter}, a free place immediately follows each white
pebble -- a contradiction.
\end{proof}

The complete proof of correctness of {\sc Delete} is presented next.

\begin{proofof}{Theorem~\ref{theorem:deleteCorrect}}
Let us number the iterations of the {\bf while} loop by $t=1,2,\ldots$.  Let
$\Gamma_t$ be the configuration of pebbles and free places at the beginning of
the $t$-th iteration.  Hence, the $t$-th iteration starts with $\Gamma_t$, and
a position $i_t$; it selects a pebble $X_t$ of size $x_t$ starting at $j_t$,
moves it to $i_t$, and sets $i_{t+1}:=j_t$. Moreover, at the beginning of the
$t$-th iteration we shall consider a free place $P_t$ of size $a_t$ starting at
$i_t$, such that $a_1=a$, and the size $a_{t+1}$ is determined as follows: let
$\Gamma'_t$ be obtained from $\Gamma_t$ by putting a new pebble $A_t$ on $P_t$.
If the color of $X_t$ in  $\Gamma'_t$ is black then let $a_{t+1}=x_t$,
otherwise let $Y_t$ be $X_t$'s left neighbor, and $a_{t+1}=y_t$. We soon prove
that this definition is correct, i.e. that $P_{t+1}$ is indeed free.


\noindent
We shall prove the by induction on $t$ the following claim: 

\vskip 2mm
\noindent
{\em
For every iteration $t$ of the {\bf while} loop, the corresponding $\Gamma'_t$ is a
valid situation.
}

\vskip 2mm
\noindent
The first iteration starts after removing $A$ from a valid situation, so the
claim for $t=1$ holds.

\vskip 2mm
\noindent
Consider the $t$-th iteration. The algorithm either stops or enters the next
iteration.  We prove that in the latter case $\Gamma_{t+1}'$ is valid, provided
$\Gamma_t'$ was valid.  $\Gamma_{t+1}'$ is obtained from $\Gamma_t$ by moving
$X_t$ to $i_t$, and putting $A_{t+1}$ to $P_{t+1}$. Since $\Gamma_t'$ is valid,
the place $P_{t+1}$ exists, and obviously is free (either directly or due to
Lemma~\ref{lemma:whiteSpaceAfter}).

First note that the swap operation on line \ref{alg:delete.2} is correct: since
$X_t$ was to the right of $i_t$ in a valid situation $\Gamma'_t$, the free
bandwidth before $i_t$ is at most $x_t$. Hence, swapping $X$ and $Q$ yields a
situation where $X$ follows immediately after $W$, and $Q$ starts immediately
after $X$ (on $i_t$).  Now we  distinguish two cases:

\vskip 3mm
\noindent
{\bf Case 1: $x_t<a_t$}\\[1mm]
In this case $X_t$ is white in $\Gamma'_t$ (it has a larger pebble $A_t$ to the
left), $Y_t$ is black and $y_t\ge a_t$.  Lemma~\ref{lemma:whiteKiller2} applied
to $Y_t$ and $X_t$ assures that there are neither free spaces nor white pebbles
between the closing position of size $2x_t$ and $j_t$.  However, since $y_t\ge
a_t>x_t$ it follows from Lemma~\ref{lemma:whiteKiller} applied to $X_t$ and
$Y_t$ that $A_t$ is black in $\Gamma'_t$. We argue that there are neither white
pebbles nor free places between $i_t$ and $j_t$: let $l$ be the closing position
of size $2x_t$.  If $i_t\ge l$, the result follows; otherwise, since $a_t\ge
2x_t$ there are two possibilities: either there are some black pebbles of size
$2x_t$ in $\Gamma_t'$ -- in this case it must hold that $a_t=2x_t$ and due to
Lemma~\ref{lemma:sequenceBlack} there is a continuous sequence of black pebbles
of size $2x_t$ between $i_t$ and $l$, or there are no black pebbles of size
$2x_t$ in $\Gamma_t'$. However, in the latter case it cannot be that $i_t<l$.
Hence, there is a continuous sequence of black pebbles between $i_t$ and $j_t$
in $\Gamma_t'$.

Consider now the colors of the pebbles in $\Gamma_{t+1}'$. $A_{t+1}$ has size
$y_t$ and hence is black.  All pebbles before $i_t$ in $\Gamma_t'$ (including
possible pebble $Q$ from line \ref{alg:delete.2}) retained their colors from
$\Gamma_t'$, as did the pebbles after $A_{t+1}$.  Pebble $X_t$ starts at $i_t$
and can be either black or white. Pebbles between $X_t$ and $A_{t+1}$ are black,
because they were black in $\Gamma_t'$.  Now we argue that the invariants
$\Gamma_{t+1}'$ is valid. 

\begin{itemize}

\item[\Ia:] First consider the situation when there was no swap on line
\ref{alg:delete.2}.  For pebbles before $i_t$ the free bandwidth before them
remains the same. The free bandwidth before any pebble after $A_{t+1}$ in
$\Gamma_{t+1}'$ did not increase from $\Gamma_t'$: $A_t$ was removed and
$A_{t+1}$ of size $y_t\ge a_t$ added. $X_t$ was moved to $i_t$ so the free
bandwidth couldn't increase. The remaining pebbles in $\Gamma_{t+1}'$ are the
pebbles between $X_t$ and $A_{t+1}$ (including $A_{t+1}$). They form a
continuous sequence of black pebbles of size at least $a_t$.  Since the free
bandwidth before $X_t$ in $\Gamma_t'$ was less than $x_t$, the free bandwidth
before each of these pebbles in $\Gamma_{t+1}'$ is less than $x_t+a_t-x_t$.

The possible swap on line \ref{alg:delete.2} doesn't increase the free
bandwidth before any pebble.

\item[\Ib:] Again, consider first the situation without a swap on line
\ref{alg:delete.2}.  The only possibility for two consecutive white pebbles in
$\Gamma_{t+1}'$ is that $X_t$ has a white neighbor.  However, $X_t$ is followed
by a sequence of black pebbles which is non-empty\footnote {If $A_t\neq
Y_t$, then this sequence contains at least $Y_t$. If $A_t=Y_t$, then
$a_{t+1}=a_t$ and since $A_t$ is black in $\Gamma_t'$, $A_{t+1}$ is black in
$\Gamma_{t+1}'$.}
the only way is that $X_t$ has a left neighbor $Q$ which is white, and hence
preceded by a black pebble $W$.  However, from Lemma~\ref{lemma:whiteIncreasing}
it follows that $q<x_t$, and from Lemma~\ref{lemma:whiteKiller} it follows that
$x_t\ge w$; hence $X_t$ is black.

Now we argue that the swap on line \ref{alg:delete.2} can not result in two
consecutive white pebbles. That could only happen if there is a white pebble
following $X_t$ before the swap. However, $X_t$ was the smallest pebble to the
right of $i_t$, hence it's right neighbor must be black.

\item[\Ic:] Since $X_t$ was white in $\Gamma_t'$, the pebble following $A_{t+1}$
in $\Gamma_{t+1}'$ is black. So the only way to violate \Ic in $\Gamma_{t+1}'$
is by means of $X_t$. Suppose that $X_t$ is white in $\Gamma_{t+1}'$. Then
either $A_t$ was the rightmost pebble\footnote{In case that $t=1$ or
$a_t=x_{t-1}$; see line~\ref{alg:delete.1} of
procedure {\sc Delete}. } of size $a_t$ in $\Gamma_t'$ or right neighbor of
$A_t$ in $\Gamma_t'$ is strictly bigger\footnote{In this case $a_t=x_{t-1}$
and the claim follows directly from $\Ic$ in $\Gamma_t'$.} than $a_t$.
Hence, in order to violate \Ic, $X_t$ must be black in $\Gamma_{t+1}'$, must have a white left
neighbor $Q$ which in turn has black left neighbor $W$ of size $x_t$. Then,
however, $X_t$ is swapped with $Q$. The result follows by noting that $Q$ has a
black right neighbor of size at least $a_t\ge x_t>q$ ($Y_t$ or some black pebble before it).

\end{itemize}

\vskip 3mm
\noindent
{\bf Case 2: $x_t\ge a_t$}\\[1mm]
First note that in this case, the swap on line \ref{alg:delete.2} never
happens. Indeed, if $A_t$ was white in $\Gamma_t'$, it was immediately preceded
by a black pebble which remains black also in $\Gamma_{t+1}'$. On the other
hand, if $A_t$ was black in $\Gamma_t'$ and had a white left neighbor $Q$, then
$Q$'s left neighbor $W$ was black, and $w<a_t\le x_t$.  Now consider the colors
of the pebbles in $\Gamma_{t+1}'$. Pebbles before $i_t$ have the same color as in
$\Gamma_t'$. $X_t$ can be either black or white, however, if it is white, then
$A_t$ must have been white in $\Gamma_t'$. $A_{t+1}$ has the size $x_t$ or
$y_t$ and is black. All other pebbles have the same color as in $\Gamma_t'$,
because $X_t$ was the smallest one.  Let us proceed to show that the invariants
hold in $\Gamma_{t+1}'$:

\begin{itemize}
\item[\Ia:] Clearly, the free bandwidth before pebbles to the left of $i_t$
remains the same. Since $a_{t+1}\ge x_t\ge a_t$, the invariant holds for pebbles
to the right of $X_t$. Finally, the invariant holds for $X_t$.

\item[\Ib:] Because $A_{t+1}$ is black, the only possibility for two
consecutive white pebbles in $\Gamma_{t+1}'$ is that $X_t$ is white and has a
white neighbor.  However, if $X_t$ is white in $\Gamma_{t+1}'$ then $A_t$ was
white in $\Gamma_t'$, and had two black neighbors. Let $L$ be $A_t$'s left
neighbor from $\Gamma_t'$. Since $X_t$ is white in $\Gamma_{t+1}'$, due to
Lemma~\ref{lemma:whiteIncreasing} it holds $a_t<x_t<l$.  If the right neighbor
of $A_t$ in $\Gamma_t'$ was different from $X_t$, it was black and remains
black in $\Gamma_{t+1}'$.  Hence, in order to violate \Ib, let the right
neighbor of $A_t$ in $\Gamma_t'$ be $X_t$. Since $x_t<l$, it is a contradiction
with the fact that $\Gamma_t'$ was valid.  

\item[\Ic:] Since $\Gamma_t'$ was
valid, there are only two places where the \Ic could be violated in
$\Gamma_{t+1}'$:  the violating configuration must involve either $X_t$ or
$A_{t+1}$. Since $A_{t+1}$ is black in $\Gamma_{t+1}'$, either $X_t$ was black
in $\Gamma_t'$ and nothing changed, or $X_t$ was white and preceded by a black
pebble of size $x_{t+1}$, and  the invariant holds, too.

Let us suppose that the violating configuration contains $X_t$. If $X_t$ is
white in $\Gamma_{t+1}'$, then $A_t$ was white in $\Gamma_t'$ and \Ic holds. If
$X_t$ is black in $\Gamma_{t+1}'$, the violating configuration must involve a
white neighbor $Q$ of $X_t$.  However, if $Q$ is $X_t$'s right neighbor, then
$x_t=a_t$, and \Ic holds. Hence, $Q$ is $X_t$'s left neighbor, and following
the argument at the beginning of Case 2, \Ic cannot be violated.
\end{itemize}

Now that we have proved the claim, let us consider the last iteration of the
algorithm. At the beginning, the statement of the claim holds. We shall argue
that at the end of the algorithm invariants \Ia--\Ic hold. Let $\Gamma_t$ be
the final configuration.

\myfig{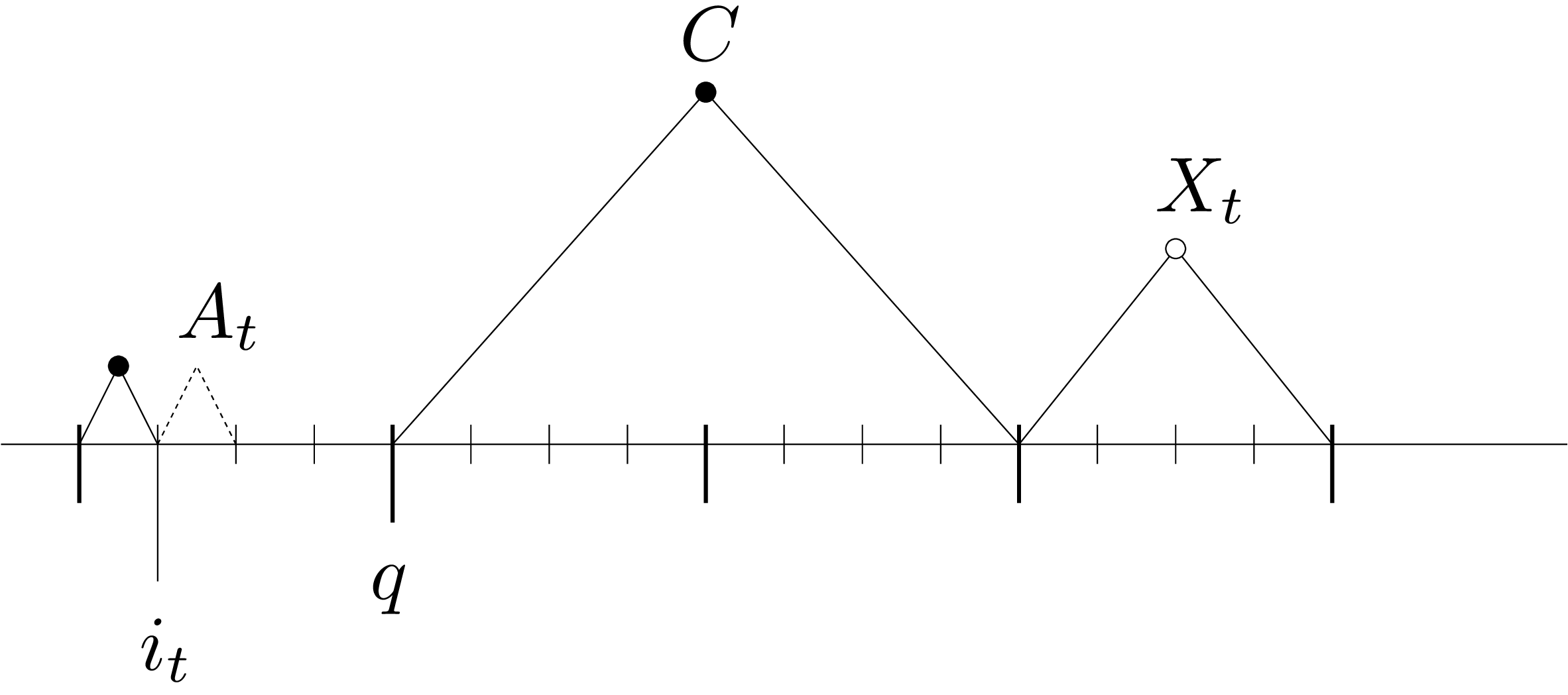}{9cm}{A possible situation in the last iteration of the {\bf while} loop.}

There are two reasons for the algorithm to stop. One of them is if, at the
beginning of an iteration, there are no pebbles to the right of $i_t$. Then
$\Gamma_t$ is obtained from $\Gamma_t'$ by deleting the rightmost pebble $A_t$
so all invariants hold. The other possibility to stop the algorithm is when the
selected pebble $X_t$ does not fit to position $i_t$.  Let $C$ be the first
pebble after $i_t$ in $\Gamma_t$ (see Figure~\ref{fig:014}).  As $x_t\ge
2a_t$ we can argue that there is no free place between $C$ and $X_t$ in
$\Gamma_t'$ (and hence in $\Gamma_t$).  Indeed, if $X_t$ is black in
$\Gamma_t'$, it holds $x_t\ge c$, but $x_t\le c$ since $X_t$ was the smallest
pebble.  Hence $x_t=c$ and due to Lemma~\ref{lemma:sequenceBlack} there is no
free place between $C$ and $X_t$. On the other hand, if $X_t$ is white in
$\Gamma_t'$, the claim holds due to Lemma~\ref{lemma:whiteKiller2}.

Moreover, the colors of the pebbles in $\Gamma_t$ and $\Gamma_t'$ are the same,
as can be seen by considering the removal of $A_t$ from $\Gamma_t'$. Pebbles to
the left of $A_t$ don't change color. Pebbles after $A_t$ are all bigger than
$A_t$ ($X_t$ was smallest of them and did not fit into $A_t$'s place), so
deleting a smaller pebble before them cannot change their color.
Now let us prove that the invariants hold in $\Gamma_t$.

\begin{itemize}
\item[\Ia:] Let $q$ be the position of $C$; we shall argue that $q-i_t\le
x_t-a_t$. Obviously, $q$ is a multiple of $x_t$. Since $\Gamma_t'$ is valid,
$A_t$ is fully contained in the place of size $x_t$ ending at $q$.  Now suppose
for the sake of contradiction that $q-i_t>x_t-a_t$. Since $i_t$ is a multiple
  of $a_t$, it must hold that $q-i_t\ge x_t$.  But then, $X_t$ would fit at $i_t$
  -- a contradiction.

The free bandwidth before $i_t$ in both $\Gamma_t$ and $\Gamma_t'$ is at most
$a_t-1$.  From the previous claim it follows that the free bandwidth before any
pebble located after $i_j$ in $\Gamma_t$ is at most $a_t-1+x_t-a_t=x_t-1$. Since
all the considered pebbles are of size at least $x_t$, the invariant holds.
\item[\Ib:] Since $C$ is bigger than $A_t$ and $\Gamma_t'$ is valid, $C$ is
black in both $\Gamma_t$ and $\Gamma_t'$. Hence, the invariant could not be
violated by deleting $A_t$.

\item[\Ic:] Since $C$ is black, the only way to violate \Ic is if $A_t$'s 
left neighbor, $W$, is white and has black left neighbor of size $c$. However,
then both $W$ and $A_t$ would have been white in $\Gamma_t'$ -- a contradiction.
\end{itemize}
\end{proofof}

As a last part in this appendix, we present the full proof of Lemma~\ref{lemma:deleteComplexity}
concerning the complexity of {\sc Delete}:

\begin{proofof}{Lemma~\ref{lemma:deleteComplexity}}
First, we treat a special case when no iterations were performed, i.e. the
rightmost pebble $A$ of given size $a$, starting at $i$, was removed, and either
there are no pebbles to the right of $A$, or the smallest such pebble $X$ does
not fit at $i$.  In this case the situation is valid due to
Theorem~\ref{theorem:deleteCorrect}, and what remains is to show how to
maintain \Id. If $A$ was the rightmost pebble, the free place formed by removing
$A$ have no pebbles to the right.  For all other free places, the size of the
smallest pebble to their right could not decrease, from which it follows that \Id
holds. If $X$ does not fit at $i$, it must be that all pebbles to the right of
$A$ are bigger than $A$. Hence, it is sufficient to put one coin at $A$, and
using similar arguments we get that \Id holds.

For the rest of the proof suppose that there is at least one full iteration of
the main loop.  Recall the notation from the proof of
Theorem~\ref{theorem:deleteCorrect}, i.e. we number the iterations of the main
loop, and $\Gamma_t$ is the configuration at the beginning of $t$-th iteration.
Moreover, $\Gamma_t'$ is obtained from $\Gamma_t$ by purring a new pebble $A_t$
on $P_t$. From the proof of the theorem it follows that $\Gamma'_t$ is always a
valid situation. We prove by induction on $t$ that the following can be
maintained:

\vskip 2mm
\noindent
{\em
For every iteration $t>1$ of the {\bf while} loop, the corresponding
$\Gamma'_t$ satisfies \Ia--\Id, all pebbles to the right of $i_t$ have size at least
$a_t$, and one extra coin lays on $A_t$. Moreover, if some
pebble to the right of $i_t$ has the size $a_t$, then two extra coins lay on
$A_t$.
}

\vskip 2mm
\noindent
{\bf Basis:} Consider $\Gamma_2'$. It is obtained from the initial situation
($\Gamma_1'$) by removing $A_1=A$, moving $X_1$ to $i_1$ with an optional swap
with its left neighbor, and placing $A_2$ on $i_2=j_1$.  We prove that all
pebbles to the right of $i_2$ have size at least $a_2$.  $X_1$ was selected as the
smallest and rightmost pebble to the right of $i_1$, hence all pebbles to the
right of $j_1$ are bigger than $x_1$. If $X_1$ was black in $\Gamma_1$,
$a_2=x_1$, so let us suppose that $X_1$ was white, and $a_2=y_1$ (see
Figure~\ref{fig:013}). Then by Lemma~\ref{lemma:whiteKiller} all pebbles to the
right of $j_1$ are of size at least $y_1$.

Let $C$ be $A_1$'s right neighbor in $\Gamma_1$, and $q$ be its starting
position (see Figure~\ref{fig:015}). Note that there is no free bandwidth
between $q$ and $j_1$ in $\Gamma_1$: Since $X_1$ is the smallest pebble to the
right of $i_1$, $c\ge x_1$, and $q$ is a multiple of $x_1$, and so are the
sizes of all pebbles between $C$ and $X_1$. Obviously, there is no free place
between $C$ and $X_1$, because otherwise there would be at least $x_1$ free
bandwidth before $X_1$ -- a contradiction with \Ia.

\myfig{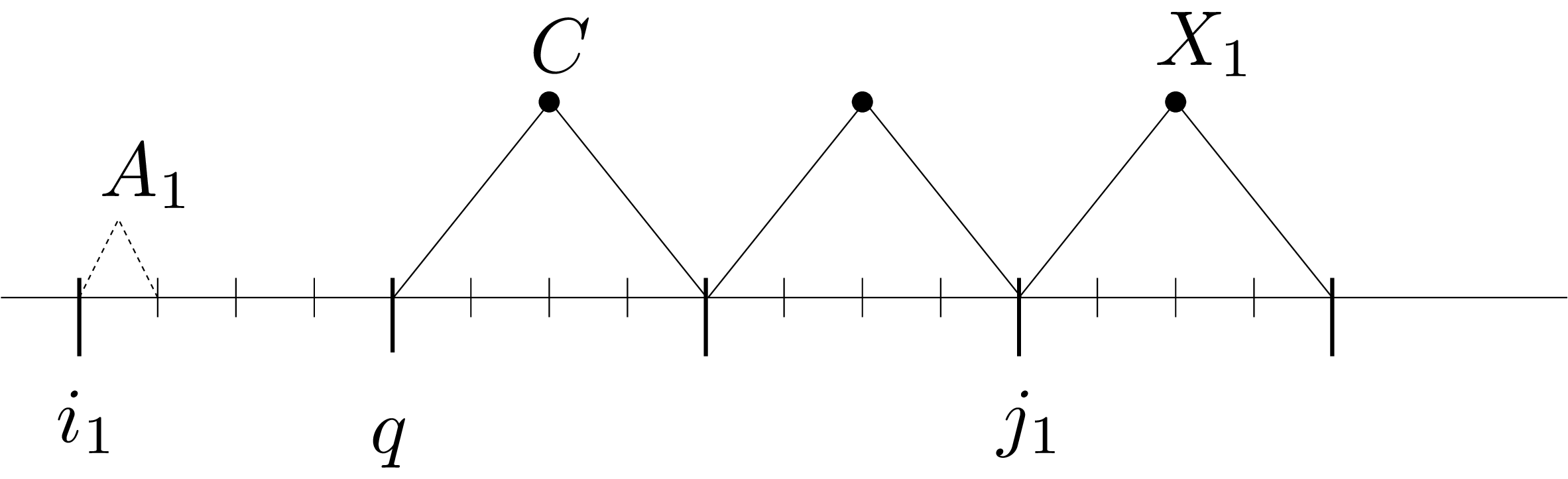}{9cm}{There is no free bandwidth between $q$ and $j_1$.}

What remains to be proven is that it is possible to maintain \Id in
$\Gamma_2'$; this is sufficient because the one or two extra coins to be put of $A_2$,
and the coin needed for the first iteration can be both charged to the constant
number of coins allowed for the {\sc Delete}.

\noindent
Let us now distinguish two cases:

\vskip 2mm
\noindent
{\em Case 1: $x_1\ge a_1$.}
In this case the swap on line \ref{alg:delete.2} cannot happen (see Case 2 in
the proof of Theorem \ref{theorem:deleteCorrect}), so $\Gamma_2'$ is obtained by
removing $A_1$, moving $X_1$ to $i_1$, and placing $A_2$ on $j_1$.  Obviously,
for all places to the left of $i_1$, \Id remains valid, since the smallest
pebble to the right could not decrease.  Also, places to the right of $j_1$ were
not affected. Hence, we can restrict ourselves to places between $i_1$ and
$j_1$.  However, there is no free place between $i_1$ and $j_1$ in $\Gamma_2'$:
there is no free place between $q$ and $j_1$ in $\Gamma_1'$ (and thus also in
$\Gamma_2'$). Moreover, since $a_1\le x_1$, $i_1$ is a multiple of $x_1$ ($X_1$
fits at $i_1$), and due to \Ia, it holds that $q-i_1=x_1$, and there is no free
place between $i_1$ and $j_1$ in $\Gamma_2'$.

\vskip 2mm
\noindent
{\em Case 2: $x_1<a_1$.}
First, consider the case without the swap on line \ref{alg:delete.2}.  As
above, it is sufficient to prove that \Id holds for free places between $i_1$
and $j_1$.  Note that in $\Gamma_1'$ there is no free place between $i_1$ and
$j_1$: if there would be a free place, it would have to be between $A_1$ and
$C$, which would, in turn, lead to contradiction with \Ia applied to $X_1$.
Hence, the only free places of interest are those remained from $P_1$ after
placing $X_1$ on $i_1$.

Since $x_1<a_1$, $X_1$ is white in $\Gamma_1'$, and $a_2=y_1\ge a_1>x_1$. We prove
that all pebbles to the right of $i_1$ in $\Gamma_2'$ (except $X_1$) are of size
at least $a_1$. We already know that all pebbles to the right of $i_2$ have
size at least $a_2 \ge a_1$. Since $X_1$ was selected as the smallest one, all pebbles
between $i_1$ and $j_1$ in $\Gamma_1'$ are of size at least $x_1$. However, due
to Lemma~\ref{lemma:whiteIncreasing} they must be black, and thus of size at least
$a_1$. The same pebbles are in $\Gamma_2'$.

Consider any free place $P$ in $\Gamma_2'$ created after the removal of $A_1$.
Since the smallest pebble to the right of $P$ is of size at least $a_1$, if
$p<a_1/2$, no coins need to be placed on $P$. However, $p\le a_1/2$, and there
is at most one $P$ for which $p=a_1/2$. Hence, one coin placed on $P$ is
sufficient to maintain \Id.

\myfig{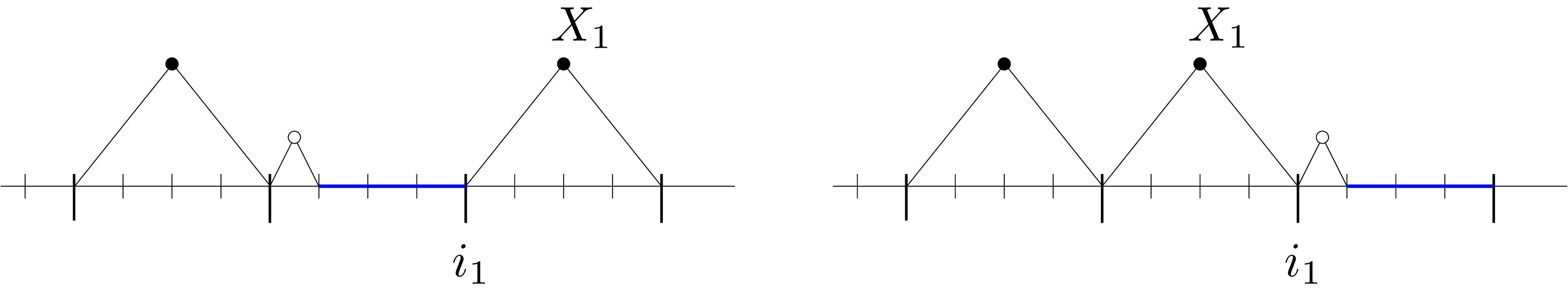}{13cm}{Moving free places during swap.}

The last thing to note is how to handle the swap on line \ref{alg:delete.2}. As
can be seen from Figure~\ref{fig:016}, the only free places that were affected
by the swap are those immediately to the left of $i_1$. However, the coins may
be moved to corresponding places; obviously, the smallest pebble to the right
cannot decrease, so we conclude that the new allocation of coins satisfies \Id.

\vskip 2mm
\noindent

{\bf Induction step:} In order to prove the claim for $\Gamma_{t+1}'$, consider
the situation when the algorithm finishes the $t$-th iteration and enters the
$t+1$st.  $\Gamma_{t+1}'$ is obtained from $\Gamma_t'$ by removing $A_t$,
moving $X_t$ to $i_t$, and placing $A_{t+1}$ on $i_{t+1}=j_t$. Note that in
this case $x_t\ge a_t$, so the is no swap.  Using the same arguments as above we
argue that all pebbles to the right of $i_{t+1}$ have size at least $a_{t+1}$, and
that there is no free place between $i_t$ and $j_t$ in $\Gamma_{t+1}'$. Since
for the free places before $i_t$ and after $j_t$, \Id remains valid, we argue
that \Id folds in $\Gamma_{t+1}'$.

Now we show how to find two free coins -- one to pay for this
iteration, and one to be placed on $A_{t+1}$. If $x_t=a_t$, there are two
coins placed on $A_t$. Otherwise (i.e. in case that $x_t>a_t$) one coin comes from the deletion
of $A_t$ and the other can be found as follows. Since $X_t$ was placed on $i_t$,
and $x_t>a_t$, there must have been a free place $P$ of size $x_t/2$ in
$\Gamma_t'$. Moreover, $X_t$ was to the right of $P$, and so there must have
been at least one coin on $P$. In $\Gamma_{t+1}'$, $P$ is covered by $X_t$, so
the coin can be used.

The last thing to show is to find a second free coin in case that there
exists some pebble $Q$ to the right of $A_{t+1}$ such that $q=a_{t+1}$. In
this case $X_t$ is white in $\Gamma_t'$ -- otherwise it would hold that
$a_{t+1}=x_t$ and there would be no pebble of size $x_t$ to the right of $X_t$.
Hence $x_t\le a_{t+1}/2$ and $A_{t+1}$ in $\Gamma_{t+1}'$ covers a place of
size $a_{t+1}/2$ that has been free in $\Gamma_t'$. According to \Id 
there is a coin on this place in $\Gamma_t'$; this coin can be used.

The proof of the theorem is concluded by considering the last iteration. Let
$\Gamma_{t_{fin}}'$ be the last situation. The final situation is obtained from
$\Gamma_{t_{fin}}'$ by removing $A_{t_{fin}}$.  We show that \Id holds. The
only free place that could violate \Id is the one remained after $A_{t_{fin}}$,
however, there was a coin on $A_{t_{fin}}$, and all pebbles to the right (if
any) are bigger that $a_{t_{fin}}$.
\end{proofof}

\end{document}